\documentclass[twoside,preprint2]{aastex}

\usepackage{graphics}
\usepackage{psfig}
\usepackage{latexsym,amssymb}

\newcommand{\teff}{$T_{\mathrm{eff}}$}
\newcommand{\mteff}{T_{\mathrm{eff}}}
\newcommand{\logg}{$\log{g}$}

\newcommand{\metal}{[M/H]}

\shorttitle{A new high-resolution theoretical library of ultraviolet spectra}
\shortauthors{Rodriguez-Merino et al.}

\begin{document} 
\title{UVBLUE: a new high-resolution theoretical library of\\
 ultraviolet stellar spectra}

\author{L.\ H.\ Rodriguez-Merino, M.\ Chavez\altaffilmark{1}, E.\ Bertone}
\affil{INAOE, Luis Enrique Erro 1, 72840, Tonantzintla, Puebla, Mexico}
\email{lino@inaoep.mx, mchavez@inaoep.mx, ebertone@inaoep.mx} 
\and \author{A. Buzzoni} \affil{INAF - Osservatorio Astronomico di Bologna,
 Via Ranzani 1, 40127 Bologna, Italy}
\email{buzzoni@bo.astro.it}

\altaffiltext{1}{Visiting Astronomer at the Steward Observatory and Lunar and
Planetary Laboratory, University of Arizona Tucson, AZ 85721.}

\begin{abstract} We present an extended ultraviolet-blue (850-4700~\AA)
library of theoretical stellar spectral energy distributions (SEDs) computed 
at high resolution, $\lambda/\Delta \lambda$= 50\,000.  The {\sc Uvblue} grid, 
as we named the library, is based on LTE calculations carried out with 
{\sc Atlas9} and {\sc Synthe} codes developed by R.~L.~Kurucz and consists 
of nearly 1800 entries that cover a large volume of the parameter space. 
It spans a range in \teff$\;$ from 3000 to 50\,000~K, the surface gravity 
ranges from $\log{g}$= 0.0 to 5.0 with $\Delta\log{g}$= 0.5~dex, while seven 
chemical compositions are considered: \metal$= -2.0, -1.5, -1.0, -0.5, +0.0, 
+0.3$ and +0.5~dex. For its coverage across the H-R diagram, this library is 
the most comprehensive one ever computed at high resolution in the 
short-wavelength spectral range, and useful application can be foreseen both 
for the study of single stars and in population synthesis models of galaxies 
and other stellar systems.

We briefly discuss some relevant issues for a safe application of the 
theoretical output to ultraviolet observations, and a comparison of our LTE 
models with the NLTE ones from the {\sc Tlusty} code is also carried out. 
NLTE spectra are found, in average, to be slightly ``redder'' compared to the 
LTE ones for the same value of \teff, while a larger difference could be 
detected for weak lines, that are nearly wiped out by the enhanced core 
emission component in case of NLTE atmospheres. These effects seem to magnify 
at low metallicity (typically \metal$\lesssim -1$).

A match with a working sample of 111 stars from the IUE atlas, with available 
atmosphere parameters from the literature, shows that {\sc Uvblue} models 
provide an accurate description of the main mid- and low-resolution spectral 
features for stars along the whole sequence from the B to $\sim$G5 type. 
The comparison sensibly degrades for later spectral types, with supergiant 
stars that are in general more poorly reproduced than dwarfs. As a possible 
explanation of this overall trend, we could partly invoke the uncertainty in 
the input atmosphere parameters to compute the theoretical spectra. In addition, 
one should also consider the important contamination of the IUE stellar 
sample, where the presence of binary and variable stars certainly works in the 
sense of artificially worsening the match between theory and observations. 
\end{abstract}



\section{Introduction}

The study of the ultraviolet (UV) properties of stars has been an expanding 
field of research in the last few decades, following the pioneering space 
missions of {\it Voyager}, {\it OAO}, and {\it IUE} satellites, in the 70's.
In the more recent years, the two  Astro missions with the {\it Hopkins 
Ultraviolet Telescope} (HUT) and the {\it Goddard High-Resolution Spectrograph} 
(GHRS), on board the {\it Hubble Space Telescope} (HST), opened the way to 
new-generation instrumentation for space observation at short wavelengths.
The {\it Space Telescope Imaging Spectrograph} (STIS), that replaced GHRS, 
was carrying out high-resolution spectra in the near-UV range (1900--3300~\AA),
while the {\it Far Ultraviolet Spectroscopic Explorer} (FUSE) has pushed the 
far-UV (912--1900~\AA) observations down to the Lyman limit with an 
unprecedented resolution \citep[$R=\,\lambda /\Delta \lambda\approx 20\,000$;]
[]{Moos00}. On the side of UV imaging, a fresh wealth of data is coming from 
the GALEX space mission, that collects observations of stars and galaxies in 
two adjacent photometric bands over the wavelength range 1350--2800~\AA.

To cope with these striking advances in the observational capabilities, we now 
urge improved theoretical tools to carry out the analysis of stellar spectra 
at comparable accuracy and resolution (see, on this line, an early exploratory 
discussion of \citealp{leck90}). In addition, a homogeneous and complete 
collection of synthetic spectral energy distributions of stars is also a 
basic demand for population synthesis codes (cf., e.g. \citealp*{b89,bcf94,w94,
lei99,vz99,bch03,rg04}), so extensively used to model the UV emission of 
high-redshift galaxies, as seen at optical wavelengths with ground telescopes 
\citep*{pet02,b03}. 

A notable, but forcedly palliative, input in this sense has come from the
empirical compilations of UV spectra, the most comprehensive one probably 
being the \citet*{wu83} stellar atlas, based on the original IUE archive 
and covering the range 1200--3300~\AA. Shortward of Ly$\alpha$, \citet*{pe02} 
also presented an important database of galactic OB-type stars observed with 
FUSE, while \citet*{w02} extended this library to the metal poor range 
including stars of the Magellanic Clouds. 

Like for the main optical data collections (e.g.\ the {\sc Stelib} library of 
\citealp*{stelib}, or the \citealp*{gs83}, \citealp*{jhc84}, and \citealp*{pk85} 
atlases), however, a recognized drawback of UV empirical libraries is that 
observations can hardly assure any homogeneous and complete coverage of the 
stellar parameters, quite often  with non-negligible discrepancies in the 
determination of the distinctive atmosphere parameters to label target stars 
(see, as an illuminating example, the compilation of chemical abundances as 
reported by \citealp*{Cay97}).

In this respect, theoretical libraries could play a key role as model 
atmospheres can be computed for virtually any desired combination of 
parameters. Moreover, thanks to more sophisticated numerical techniques and 
computer performances, calculations can now rely on complete and accurate 
opacities, leading to a  more refined treatment of line blanketing (see, e.g.
\citealp*{Lanz03}), also accounting for non-standard physical conditions such 
as relaxed hydrostatic or thermodynamical equilibrium \citep*{Hau95,Schaerer97, 
gl99,Kudritzki00}.

A general weak point for most of the current theoretical datasets, however, 
is a quite poor spectral resolution,\footnote{The currently available Kurucz 
SEDs, for instance, sample the UV spectral range at a 10~\AA\ step.} and this 
poses an obvious limitation when matching the high-resolution observations now 
available. A tentative effort to overcome this gap is that of \citet*{Brown96}, 
who provided a synthetic grid of about 1500 UV spectra at an increased 
resolution of 3~\AA\ between 900 and 1800~\AA; this dataset, however, is 
especially tuned for old (low-mass) stars, with a limited coverage of stellar 
parameters and based on some time-saving approximations in the computing 
algorithm.

All previous arguments gave us a strong motivation to further
tackle the problem of the UV emission of stars and embark, in this paper, 
on a systematic computation of synthetic SEDs at high spectral resolution 
($R$=~50\,000) in this wavelength range. The {\sc Uvblue} grid, as we named our 
theoretical library, is based on the \citet*{Kurucz93a} code, and is primarily 
aimed at complementing stellar libraries in the ultraviolet-blue spectral 
range currently in use for stellar populations studies. Nevertheless, we also 
foresee a fruitful use of our output for a diagnostics of  individual stars, 
especially when optical information can be added to the UV data, and suitably 
matched with model grids at longer wavelength, in order to determine the fundamental 
parameters like effective temperature (\teff), surface gravity (\logg) and 
chemical composition (\metal) by matching observed spectra (see, e.g. \citealp*{bb04}).

We will arrange our discussion presenting, in Sec.~2, a brief summary of the
Kurucz original models atmospheres, that served as a main reference for our 
calculation. In Sec.~3 we list the main computational characteristics we have 
considered when using {\sc Synthe} spectrum synthesis codes and we present the 
resulting atlas. Section 4 is a critical overview of some physical issues
for a proper use of our theoretical library in the comparison with UV
observations. A direct test is carried out in this section, matching the
IUE low-resolution spectra of an extended sample of stars of different
spectral types. Our relevant conclusions are finally summarized in Sec.~5, 
together with some hints for future work.


\section{The Kurucz model atmospheres} 

The reference code for our atmosphere model calculations is {\sc Atlas9}, which 
has been constructed by \citet{Kurucz93a} in the so-called ``classical-model 
approximation'', namely assuming plane parallel atmosphere layers in 
hydrostatic and local thermodynamic equilibrium (LTE), and with fixed chemical 
composition.  The physical variables are constant with time; additionally, 
convective equilibrium is accounted for through the mixing-length theory 
\citep{bohm} with a mixing-length to pressure scale-height ratio of 
$\ell/H_p = 1.25$. The current version of {\sc Atlas9} is the result of a 
long-standing upgrade process of the original version of the code 
\citep*{Kurucz79}. As a major improvement over the years, special care has been
devoted to a refined treatment of UV emission in the model atmospheres, both 
in terms of better input physics and computational performances. In particular, 
three important upgrades have been introduced in the current version of the 
code.

{\it i)} Inclusion and correction of iron and iron-peak element opacities 
\citep*{Kurucz92}. Currently, the line lists constructed by \citet*{Kurucz93b}
comprises nearly 60 million lines of atoms and diatomic molecules, of which 
42 million correspond to iron group species. The line lists include up to nine 
ionization stages \citep*{Kurucz95}.

{\it ii)} Atmosphere spatial resolution has been increased to 72 layers, and 
calculations encompass a range in the Rosseland optical depth from $\log 
\tau_{\rm Ross} \sim 2$ ``outward'' to $\log \tau_{\rm Ross} \sim -7$.

{\it iii)} In addition to the standard mixing-length theory, convection could 
be accounted for in the models according to the so-called ``approximate 
overshooting'' \citep*{Castelli97}. However, while this approach is found to 
provide a better fit to the solar spectrum, it is a still controversial 
question whether it definitely holds in general also for real stars other than 
the Sun \citep{Castelli97}. For this reason we eventually relied on the standard 
mixing-length treatment for our models.

\begin{figure}[t] 
\psfig{file=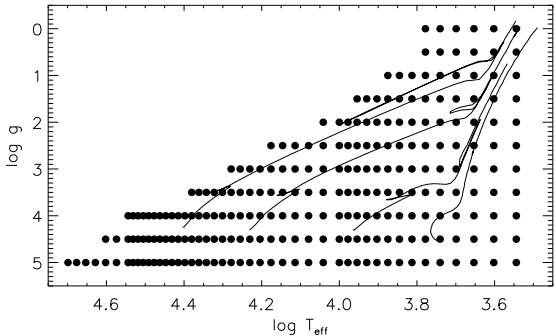,width=\hsize,clip=}
\caption{Parameter space covered by the {\sc Uvblue} grid (dots) overplotted 
with a set of solar chemical composition evolutionary tracks for 1, 2, 5 and 
10 M$_\odot$ \citep{gbbc00,salasnich}. 
\label{grid}} 
\end{figure}

For the sake of homogeneity, we decided to recompute a substantial portion of
Kurucz's grid of model atmospheres  using John Lester's modified version of 
{\sc Atlas9} adapted to run under Unix systems.\footnote{A downloable version 
of Lester's program is available at {\tt http://ccp7.dur.ac.uk/library.html}}
The parameter space covered by our library (and for which we provide a
high-resolution synthetic spectrum) is summarized in Fig.~\ref{grid}. The plot 
displays the model grid for solar metallicity, compared in the \logg~vs.\ 
log~\teff\ plane with four reference tracks reporting the evolution of stars
of 1, 2, 5 and 10 M$_{\odot}$, from \citet*{gbbc00} and \citet*{salasnich}.
One sees that effective temperature is explored over the range $\mteff= 3\,000 
\to 50\,000$~K, with step of 500, 1000 and 2500~K in the intervals 
3000--10000~K, 10000--35000~K and 35000--50000~K, respectively; surface 
gravity ranges from \logg= $0.0 \to 5.0$~dex, with steps of 0.5~dex and seven 
metallicity values are included, namely \metal= --2.0, --1.5, --1.0, --0.5, 
+0.0 (solar), +0.3 and +0.5 dex. The metal mix is solar-scaled, assuming the 
\citet*{Anders89} chemical abundances for the Sun. All models have been computed 
with a microturbulent velocity $\xi= 2$~km\,s$^{-1}$.

\section{UVBLUE: computational characteristics of synthetic SEDs} 

The whole database of stellar atmospheres accounted for in our work amounts to 
1770  models. For each of these we have computed a high-resolution synthetic 
spectrum resolving the transfer equation by means of the series of codes 
{\sc Synthe}, developed by \citet{Kurucz93b}, and also conveniently modified in 
their Unix version by John Lester.\footnote{Our calculations have been mainly 
carried out with a SUN ULTRA 10 computer at the Instituto Nacional de 
Astrofisica Optica y Electronica, Puebla, Mexico.} In our calculations we 
included all the atomic lines available in the \citet{Kurucz92} database with 
empirically determined atomic constants plus all the diatomic molecular lines 
(namely C$_2$, CN, CO, H$_2$, SiO, CH, NH, OH, MgH, and SiH) {\it excepting} 
the TiO molecule, with its nearly 37 million lines. We also excluded the 
``predicted'' lines, that is 
those lines that ostensibly generate from theoretical atomic transitions but 
are so far undetected in the observed solar spectrum (\citealp{Kurucz92}, see
also \citealp*{bert05} for a further discussion on this subject), as well as
the very weak absorption lines, that is those with residual intensity of less 
than 0.1\% with respect to the continuum. This cut is justified by the fact 
that these weak features would barely be detectable even with the highest 
reachable resolution in observed UV spectra. 

These simplifications in the {\sc Synthe} code resulted in a big saving time 
greatly speeding up the synthesis of stellar SED and allowing us to maintain 
the CPU time for the full theoretical library within reasonable limits.
Quite importantly, note that by neglecting the TiO contribution we only
very marginally affect the reliability of our output as this molecule is 
expected to become a prevailing contributor at optical/red wavelength in 
M-type stars cooler than 4000~K. At these low temperature, the UV emission of 
stars drops at nominal values (a fraction  well less than 0.1\% of the 
bolometric luminosity is emitted shortward of 3000~\AA\ by late-type stars, 
cf. \citealp*{b02}).

\begin{deluxetable}{ll}
\tablewidth{0pt}
\tablecaption{Distinctive features of the {\sc Uvblue} spectral library}
\tablehead{\colhead{Parameter}           & \colhead{Value}} 
\startdata
\teff                  &  $3000-50\,000$~K  \\
\metal                 &  $-2.0, -1.5,-1.0, -0.5,+0.0, +0.3, +0.5$ dex \\
\logg                  &  $ 0.0-5.0$ dex at steps of 0.5 dex  \\
Microturbulent velocity ($\xi$) & 2~km\,s$^{-1}$ \\    
Wavelength range       &  $ 850-4700$ \AA  \\
Resolving power, $R = \lambda/\Delta \lambda$ & 50\,000  \\
Number of spectra      & 1770  \\
Wavelength points per spectrum      & 342\,016  \\
\enddata
\label{tbl-1}
\end{deluxetable}

The main computational characteristics of {\sc Uvblue} spectral library are
summarized in Table~\ref{tbl-1}; we just recall here the two basic features.

$\bullet$) {\bf Resolving power:}  $R$= 50\,000 over the full library.  This 
high resolution represents a substantial increase with respect to previous
calculations for such a wide coverage in stellar parameters. It stands above
the resolution capabilities of most current and past UV-space missions. In
fact, the only instruments that have provided spectra of comparable resolution 
are GHRS and STIS, that reach up to $R$= 100\,000 when used with the 
echelle grating setup.

$\bullet$) {\bf Wavelength interval:} from 850 to 4700~\AA.  This interval 
encloses most of the UV range with the only exception of the extreme 
ultraviolet (EUV; 100 $\leq \lambda \leq$ 850~\AA). To ease a further match, 
we also assured a 1200~\AA\ overlap with {\sc Bluered}, the blue-optical 
spectral grid of \citet{bert05}, that self-consistently complements the 
present work at larger wavelength.\footnote{The full set of high-resolution 
SEDs of {\sc Uvblue} and the complementary optical atlas {\sc Bluered} 
by \citet{bert05} are available upon request from the authors or through the 
dedicated Web sites {\tt http://www.bo.astro.it/$\sim$eps/uvblue/uvblue.html} 
and {\tt http://www.inaoep.mx/$\sim$modelos/uvblue/uvblue.html}.}

\begin{figure*}[!ht] 
\centerline{
\psfig{file=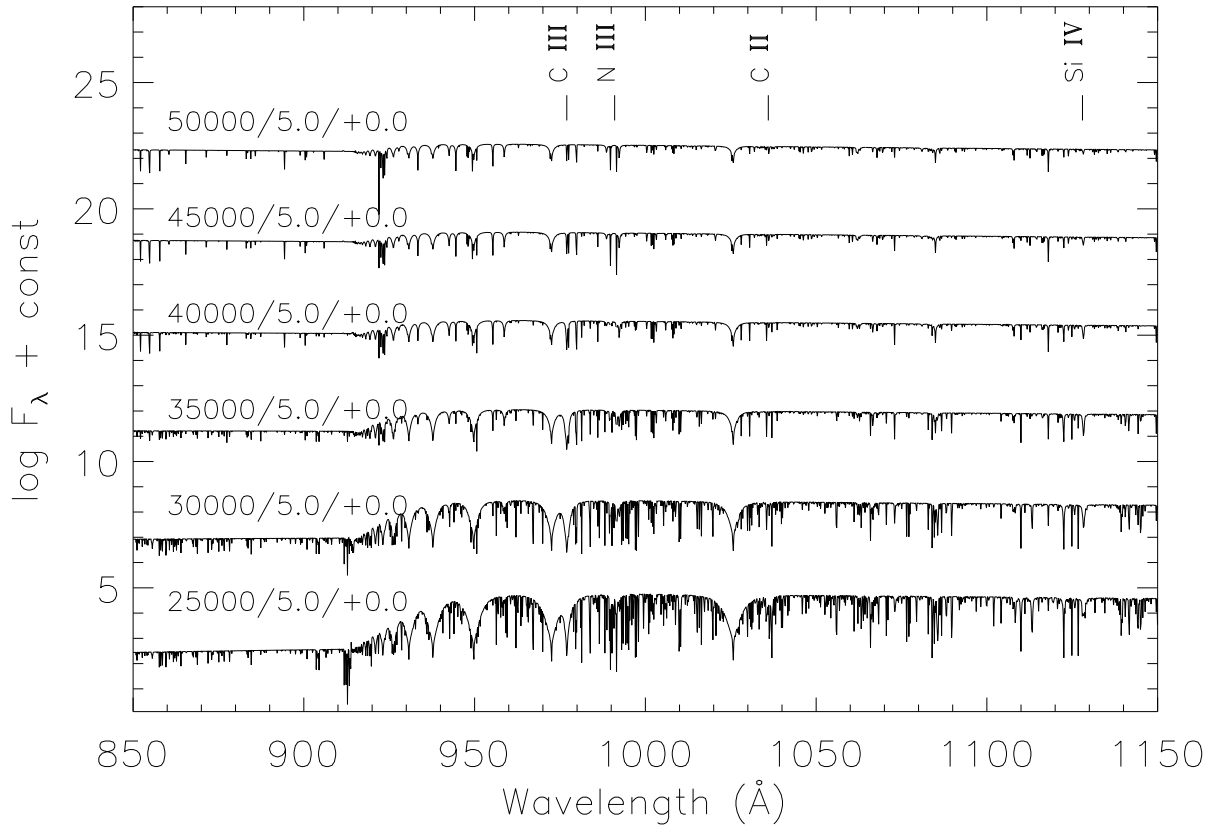,width=\hsize,clip=}
}
\caption{High-resolution ($R$= 50\,000) synthetic spectra around the 912~\AA\ 
Lyman break region [with sequence notation: (\teff/\logg/\metal), as labeled 
for each plot]. As an illustrative example, the figure explores the trend of 
the break vs.\ \teff\ for hot O-B stars of solar metallicity and \logg= 5~dex. 
A number of other interesting features, such as those of C~\textsc{iii}, 
N~\textsc{iii}, C~\textsc{ii}, and Si~\textsc{iv} at 977, 991, 1036, and 
1128~\AA, respectively, can also be recognized in the plots. [This figure has
been modified to fit the memory size contraints.]}
\label{lyman}
\end{figure*}

\begin{figure*}[!ht] 
\centerline{
\psfig{file=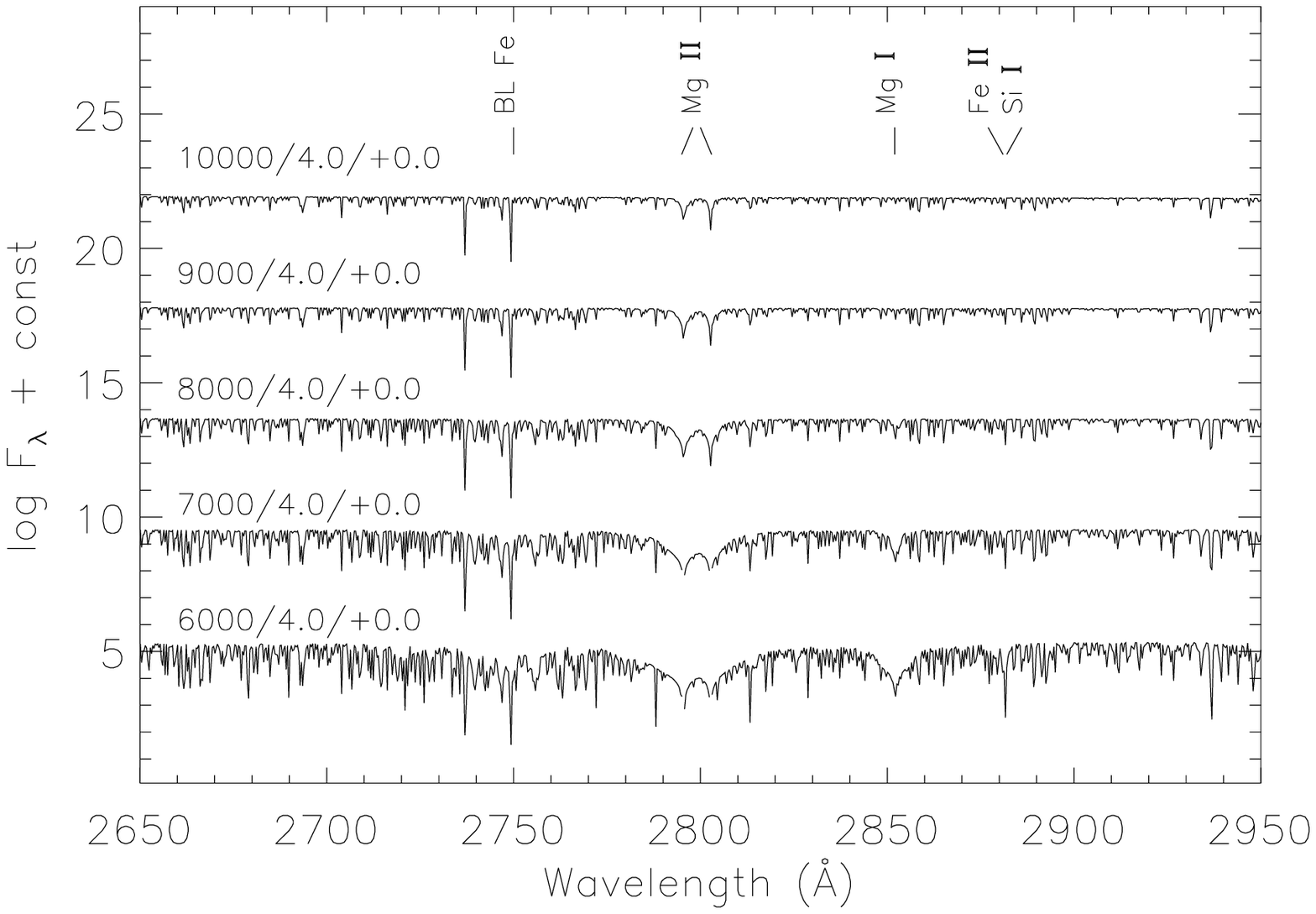,width=0.85\hsize,clip=}
}
\centerline{
\psfig{file=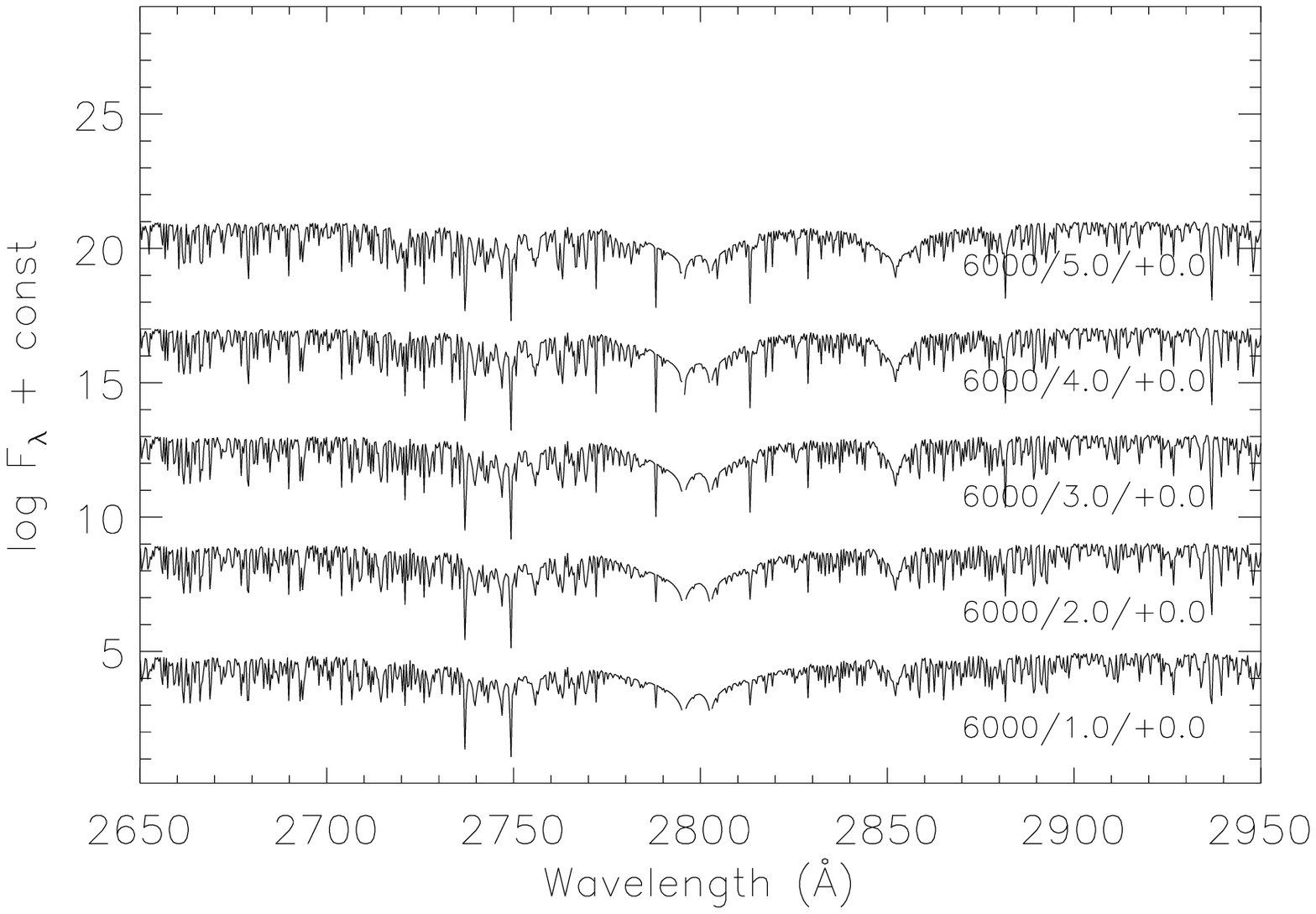,width=0.5\hsize,clip=}
\psfig{file=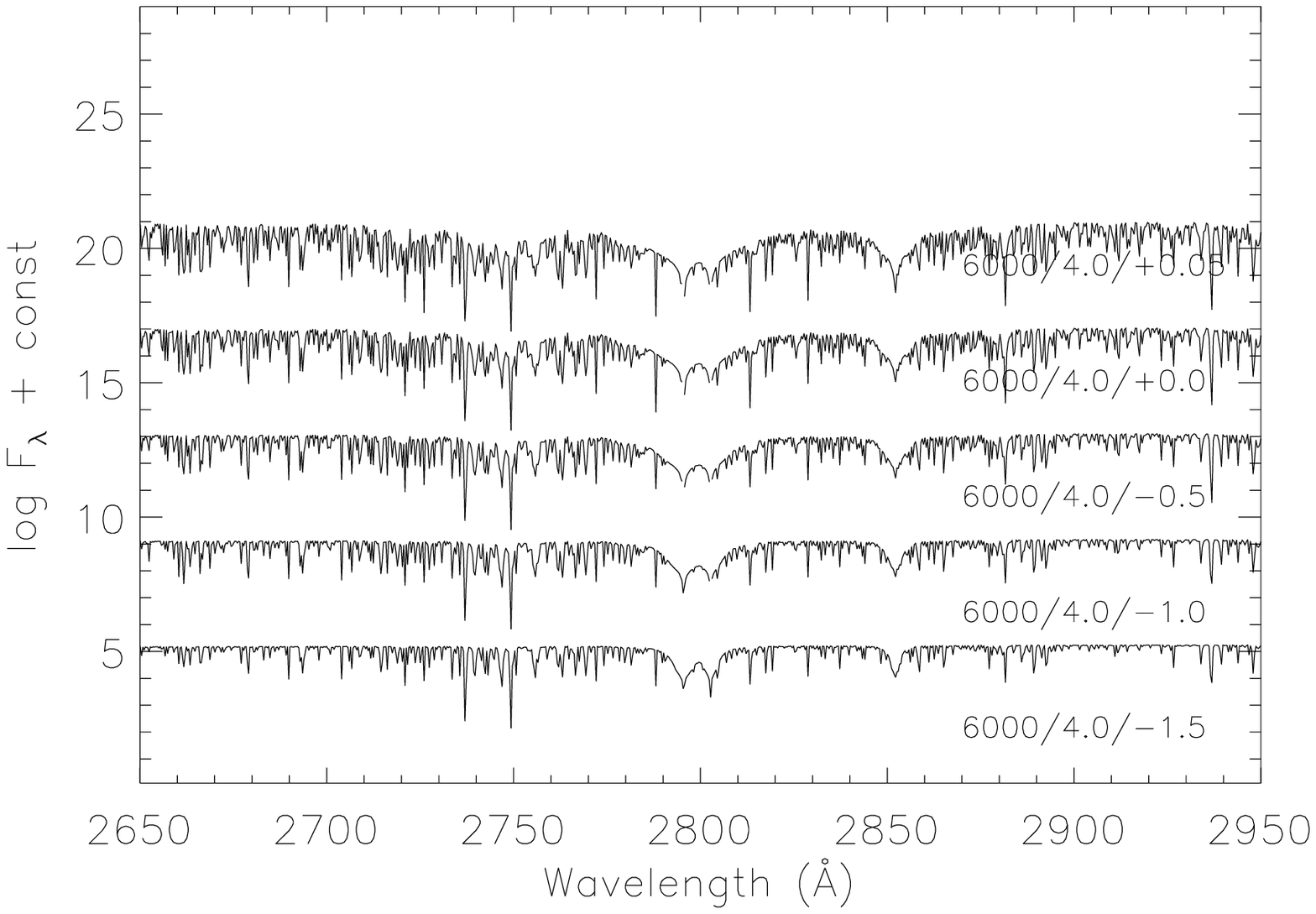,width=0.5\hsize,clip=}
}
\caption{High-resolution ($R$= 50\,000) synthetic spectra in the wavelength 
region around 2650--2950 \AA, for different choices of stellar fundamental 
parameters. The explored region includes the strong Mg~\textsc{ii} doublet at 
2795 \AA\ and 2803 \AA, very prominent in mid-F to early G-type stars and in 
early-type systems. {\it Upper panel:} the trend of SED vs.\ \teff\ for 
synthetic spectra of solar metallicity and \logg= 4.0~dex. {\it Left bottom 
panel:} same, but exploring the SED variation vs.\ surface gravity, for 
synthetic spectra with \teff= 6000~K and \metal= 0.0. {\it Right bottom panel:} 
sequence of spectra for different chemical composition, adopting \teff= 6000~K 
and \logg= 4.0~dex. [These figures have been modified to fit the memory size 
contraints.]}
\label{mg}
\end{figure*}

Figures~\ref{lyman} and \ref{mg} display some illustrative examples of the 
{\sc Uvblue} library in selected spectral regions. The 912~\AA\ Lyman break is 
explored in Fig.~\ref{lyman} for O-B stars of solar metallicity and 
\logg=~5.0~dex. On the plot we can also recognize the features of C~\textsc{iii}, 
N~\textsc{iii}, C~\textsc{ii}, and Si~\textsc{iv} at 977, 991, 1036, and 
1128~\AA, respectively. The interesting wavelength interval around the strong 
Mg~\textsc{ii} doublet at 2800~\AA\ is also displayed in the three panels of 
Fig.~\ref{mg} for different parameter sequences as indicated in each plot. 
Together with the Mg doublet, it is also evident in the figure the contribution 
of Mg~\textsc{i} features around 2852~\AA, as well as the blend of several 
Fe~\textsc{i} and Fe~\textsc{ii} features at about 2750 \AA. The Fe~\textsc{ii} 
and Si~\textsc{i} lines at 2880 and 2881~\AA $\;$ can also be 
distinguished.

\section{Comparing with the observations}

A full comparison of our theoretical output and the observed SED of real stars 
along the whole spectral-type sequence is the obvious (and mandatory) step to 
assess the robustness of our results and probe the accuracy of theoretical 
spectra in reproducing the main distinctive features as observed in stars.

Before carrying out this comparison, however, it is important to consider some 
relevant aspects that deal with the input physics of our models and should be 
kept in mind when matching observations.

\subsection{Caveats}

As a first substantial characteristics of our library, we are dealing with 
models in hydrostatic equilibrium; consequently, they should not be compared 
with observed spectra in wavelength regions that form in the winds of hot 
atmospheres. This is the case, for instance, of the strong C~\textsc{iv} line 
at 1550~\AA\ that shows a prominent P-Cygni profile in main sequence stars 
earlier than O7 or B0 in lower gravity objects.

\begin{figure}[!t] 
\psfig{file=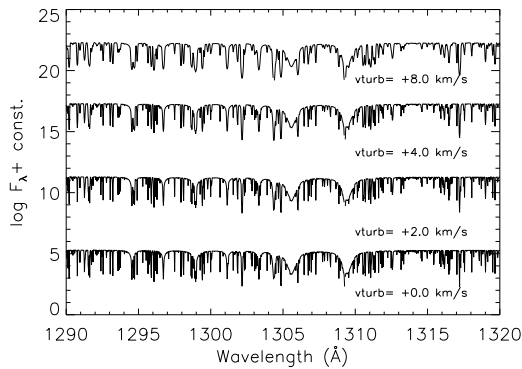,width=\hsize,clip=}
\caption{The effect of microturbulent velocity on Si/O spectral features in 
the wavelength interval 1290--1320 \AA. The same synthetic spectrum for a 
\teff= 14\,000~K star with \logg= 4.0~dex and solar metallicity is computed 
by changing the value of $\xi$, as labeled on the plots.}
\label{micro} 
\end{figure}

Second, we have only considered one value for microturbulent velocity, 
$\xi$= 2 km\,s$^{-1}$ (i.e.\ the canonical value in the Kurucz atmospheres). Non local 
thermodynamic equilibrium (NLTE) analyses of high-resolution spectra for 
early-type stars in the SMC observed with STIS and the AAT \citep*{bu03},  
indicate that values of $\xi$ as high as 10~km\,s$^{-1}$ might be more appropriate for 
the observed physical conditions. For late and intermediate-type stars various 
authors claim that more suitable values for $\xi$ are in the range 
1.0--2.5~km\,s$^{-1}$ depending on surface gravity \citep[e.g.][]{bpk03}. We plan to 
compute synthetic spectra for other values of $\xi$; however, we anticipate 
that its effects on most lines will be negligible unless very high resolution 
is considered.  To illustrate this statement, in addition to the standard case, 
we have computed in Fig.~\ref{micro} three additional models with 
microturbulent velocities of $\xi$= 0, 4, and 8~km\,s$^{-1}$. For the comparison we 
selected the region around 1300~\AA, dominated by Si~\textsc{ii}, Si~\textsc{iii} 
and O~\textsc{i} lines. The spectra are for a \teff= 14\,000~K star with 
\logg= 4.0~dex, a combination that should maximize the blend strength of 
these absorption features when seen in low-dispersion spectra \citep*{Fanelli92}.

\begin{figure}[!ht]  
\psfig{file=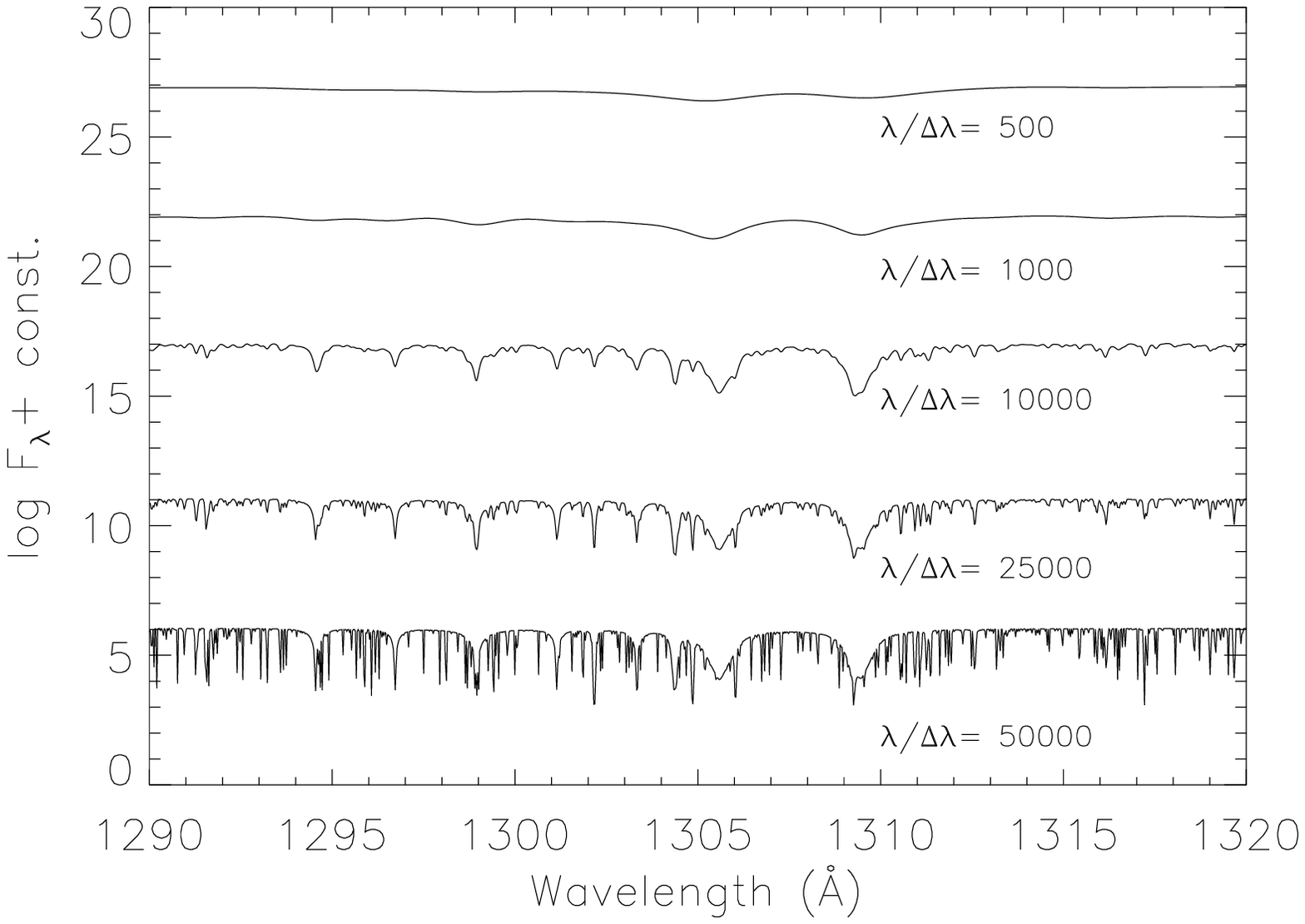,width=0.75\hsize,clip=}
\psfig{file=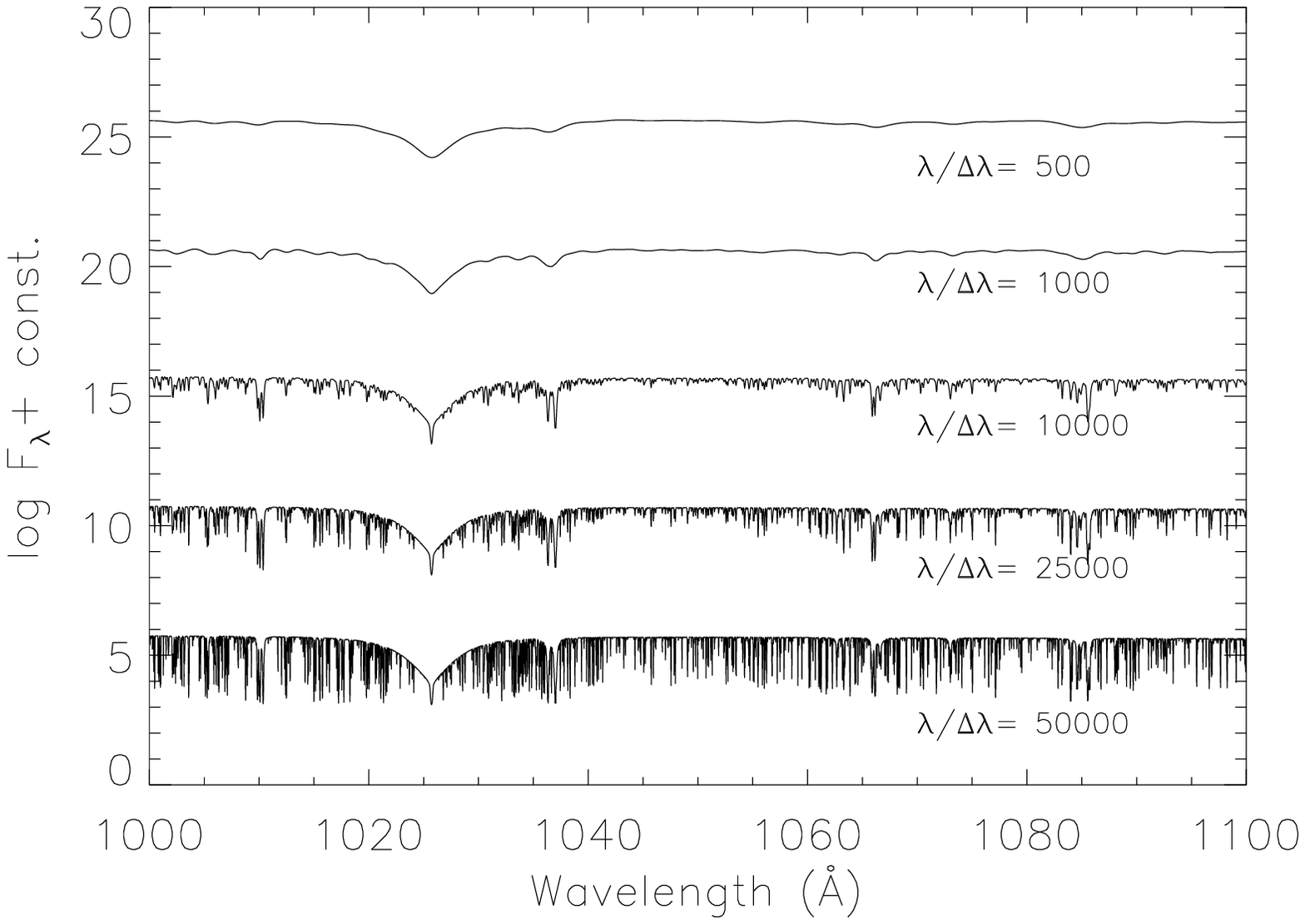,width=0.75\hsize,clip=}
\caption{An illustrative example of the effects of instrumental resolution 
($R~= \lambda/\Delta \lambda$), on a spectrum with (\teff, \logg, \metal) 
= (15000~K, 4.0~dex, +0.0~dex) for the wavelength region of Fig.~\ref{micro} 
({\it upper panel}), and for the Ly$\beta$ region about 1000--1100~\AA\ 
({\it lower panel}). The original $R$=~50\,000 spectra have been degraded by 
convolution with a Gaussian kernel.}  
\label{broad}
\end{figure}

Third, we have to make sure that spectral resolution in the grid and in
observational data are compatible.  An illustrative example is shown in 
Fig.~\ref{broad}, for the same wavelength interval of Fig.~\ref{micro} and for 
the Ly$\beta$ region about 1030~\AA. By means of Gaussian filter convolution, we 
degraded the spectra to four different resolutions, that is $R$= 500, 1000, 
10\,000, and 25\,000. Note that at the IUE low-resolution mode (i.e. $R\simeq$ 
250 at 1500~\AA) most of the features would definitely be washed out, and 
things could get even worse comparing with the nominal resolution of the 
original Kurucz grid of spectra, where $R\simeq$ 150 at 1500~\AA.

\begin{figure}[!t]  
\psfig{file=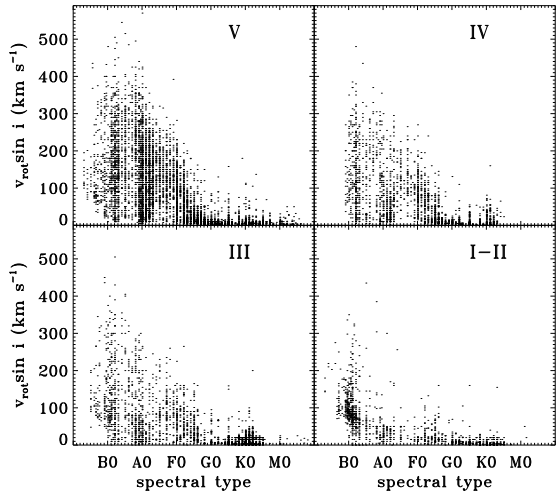,width=\hsize,clip=}
\caption{Distribution of projected rotational velocity $V \sin{i}$ vs. spectral 
type and luminosity class. Data are from the \citet{gs00} catalog, including 
only ``normal'' stars. [This figure has been modified to fit the memory size 
contraints.]}
\label{vsini}
\end{figure}

Fourth, our calculations do not include the effects of rotational velocity.
While a vast majority of late-type stars does not show any large projected 
rotational velocity, we know that most of early-type stars are severely 
affected by rotation reaching values of $V \sin{i}$ up to 500~km\,s$^{-1}$. Recently, 
\citet*{gs00} presented a compilation of over 17000 measurements of $V \sin{i}$ 
for nearly 12000 stars that complement previous catalogs \citep*{uf82}. In order 
to illustrate the overall dependence of $V \sin{i}$ on effective temperature 
and surface gravity we plot in Fig.~\ref{vsini} the $V \sin{i}$ distribution 
vs.\ spectral type for four different luminosity classes. Only the subset of 
``normal'' stars is considered in the plot, that is excluding white dwarfs, 
Wolf Rayet, binary stars etc. Evidently, the importance of rotation when 
comparing observed and {\sc Uvblue} spectra will depend on how the Doppler 
broadening compares with instrumental resolution.

Fifth, late-type stars are recognized to display some chromospheric activity
to different extent. Non radiative heating in the upper atmosphere of F to
M-type stars, through acoustic waves in F0--F5 stars and magnetohydrodynamic
mechanisms for later types (\citealp*{bohm01}, and reference therein), causes 
a turnover in the temperature stratification, $T$($\tau$). This temperature 
distribution has important effects on the SED at short wavelength as it could 
trigger prominent emission lines, like for instance the strong Mg~\textsc{ii} 
doublet at 2800~\AA\ \citep[e.g.][]{blanco}, and an overall enhancement in the 
UV continuum \citep{franchini}. Classical models in general, and Kurucz's ones 
in particular, do not account for this phenomenon since temperature decreases 
monotonically in the outermost atmosphere layers.

\subsection{LTE vs.\ NLTE}

It has been widely discussed in the literature that deviations from LTE are to 
be expected for hot stars and the cores of strong lines in stars of 
intermediate type \citep*{mihalas}. However, it has also been demonstrated that 
at low resolution the agreement between LTE and NLTE model fluxes is good, and 
that comparison of far-UV and mid-UV observational data with LTE model fluxes 
also results in good agreement \citep*{Morales00}. At high resolution such a 
comparison has been anticipated by \citet{Lanz03} but not yet quantitatively 
assessed. 

As the whole {\sc Uvblue} spectral library has been computed under the 
assumption of LTE, it may be of some interest to compare the {\sc Atlas9} 
results with the corresponding output for NLTE codes. Here, we considered in 
particular two representative sets of stellar parameters with (\teff, \logg) = 
(40\,000~K, 4.5~dex) and different metallicity, namely  \metal= 0.0~dex 
(solar) and --2.0~dex. The NLTE models and theoretical SEDs have been extracted 
from the {\sc Ostar2002} grid \citep{Lanz03}, based on line-blanketed model 
atmospheres computed with {\sc Tlusty} \citep*{Hub92} under the condition of 
hydrostatic equilibrium. Available spectral fluxes cover the wavelength region 
from 45~\AA\ to 300~$\mu$m at a resolution $R\simeq 40\,000$ and assume a 
value of $\xi = 10$~km\,s$^{-1}$. In order to make the datasets mutually consistent, 
we have recalculated the corresponding {\sc Atlas9} spectra for the same 
microturbulent velocity and wavelength resolution.

\begin{figure*}[!ht] 
\centerline{
\psfig{file=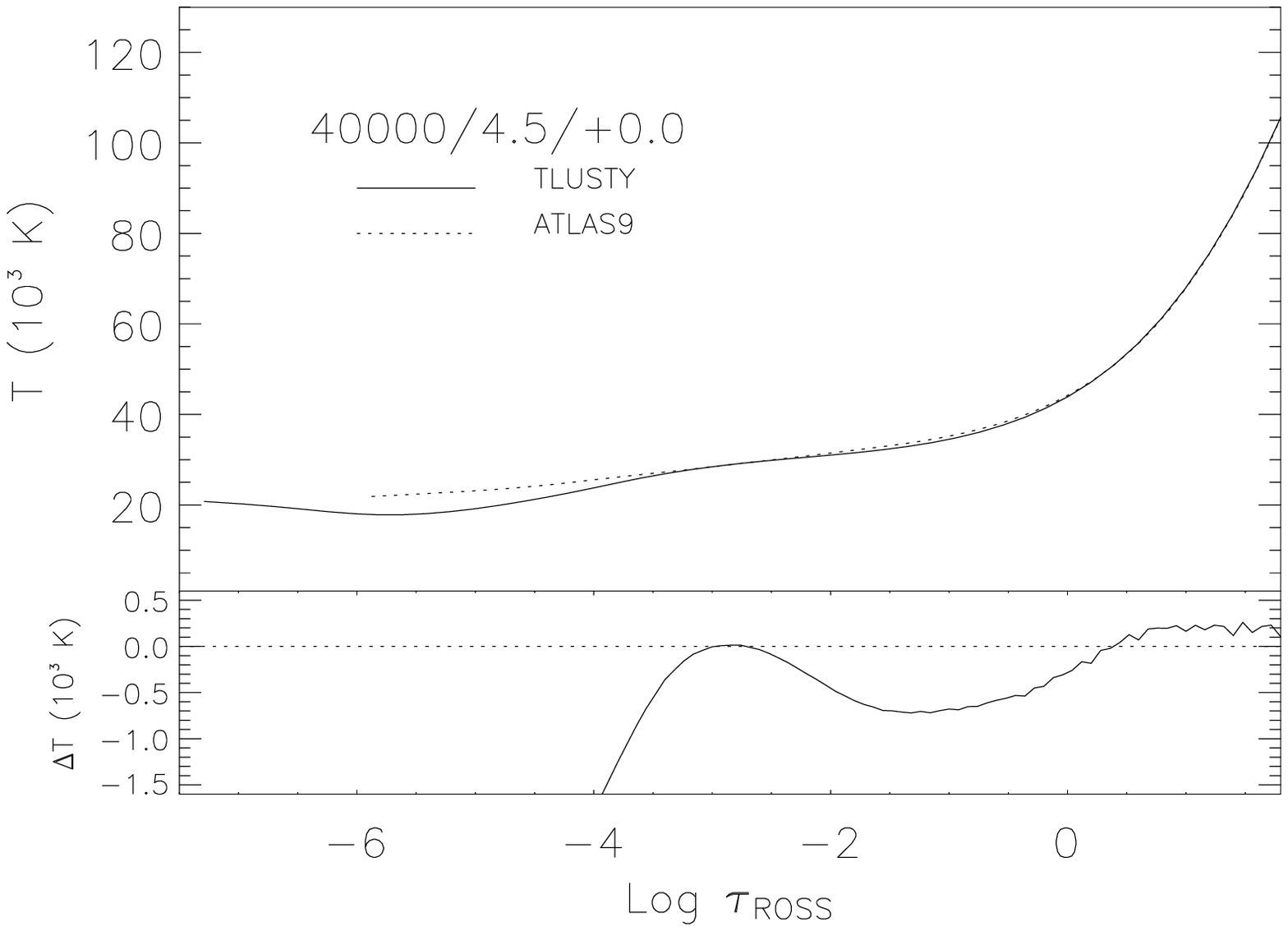,width=0.5\hsize,clip=}
\psfig{file=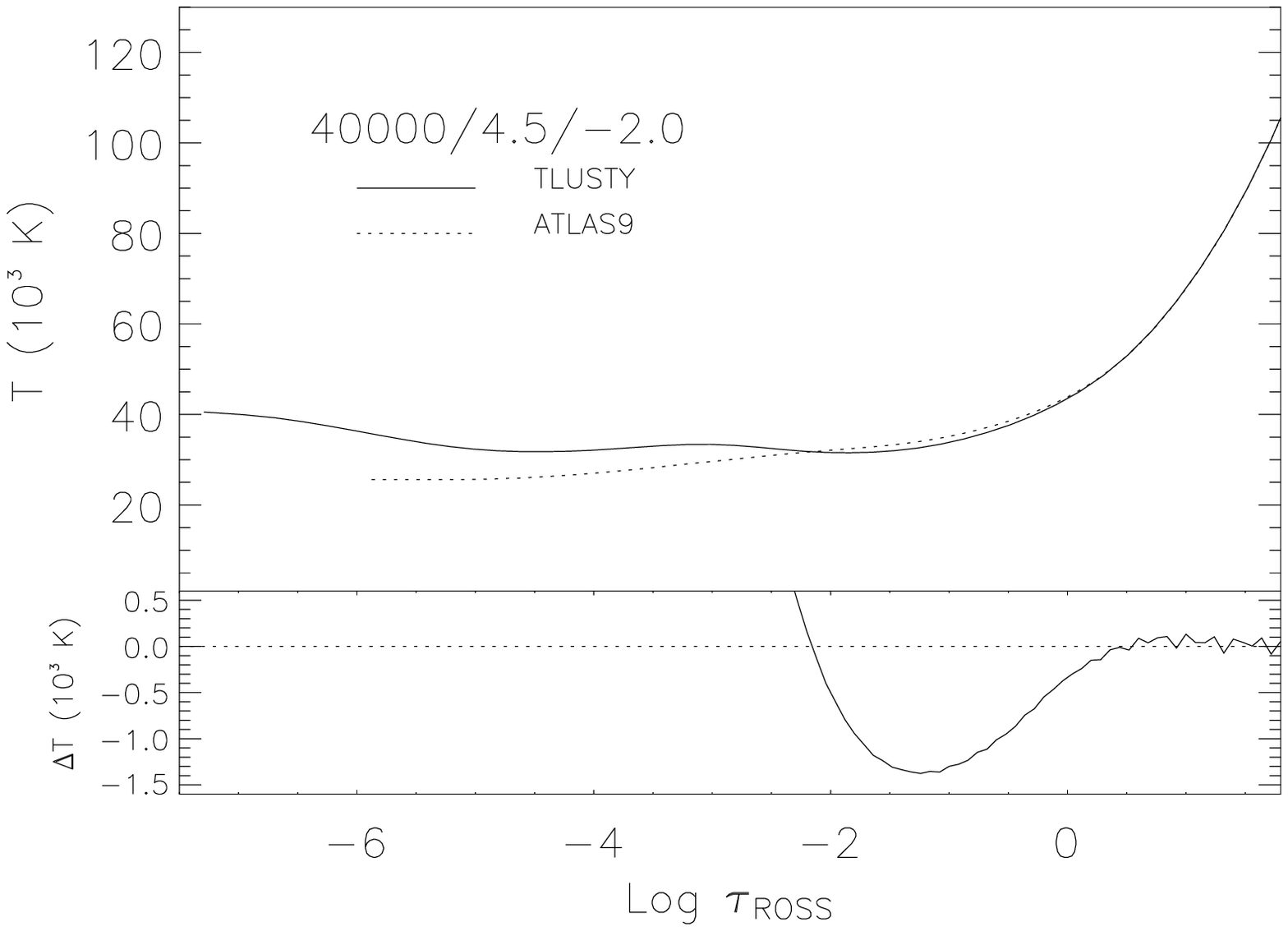,width=0.5\hsize,clip=}
}\centerline{
\psfig{file=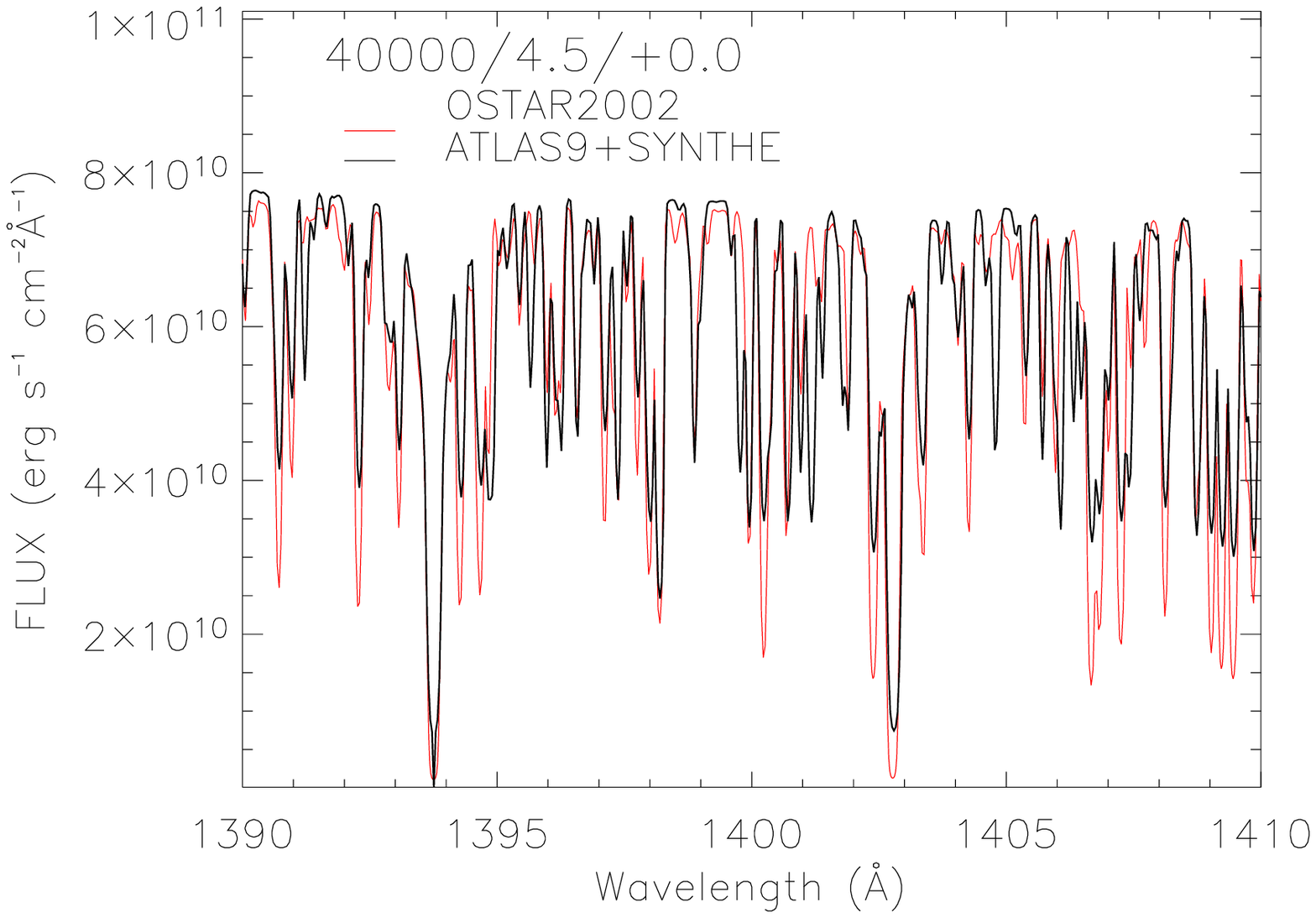,width=0.5\hsize,clip=}
\psfig{file=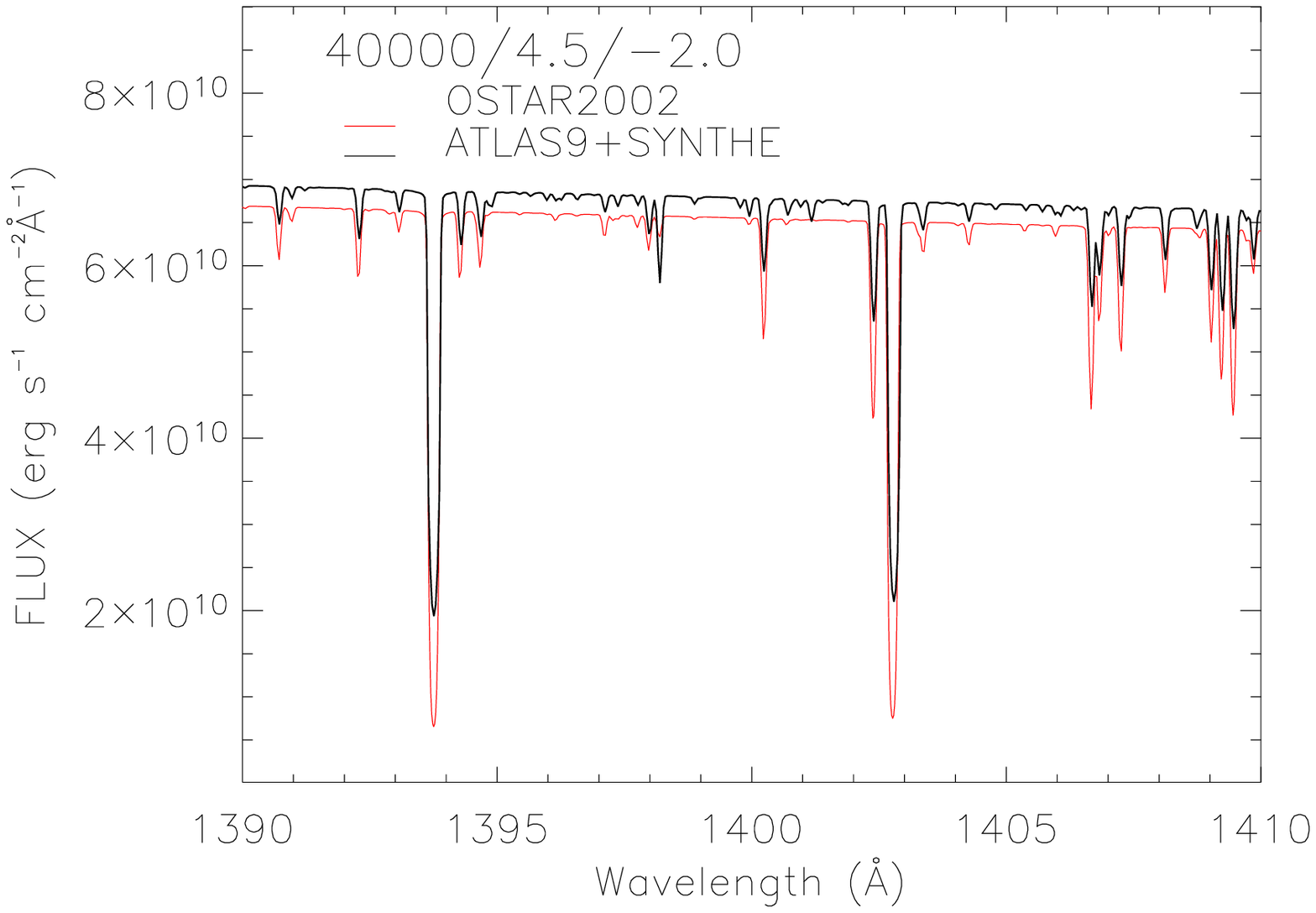,width=0.5\hsize,clip=}
}\centerline{
\psfig{file=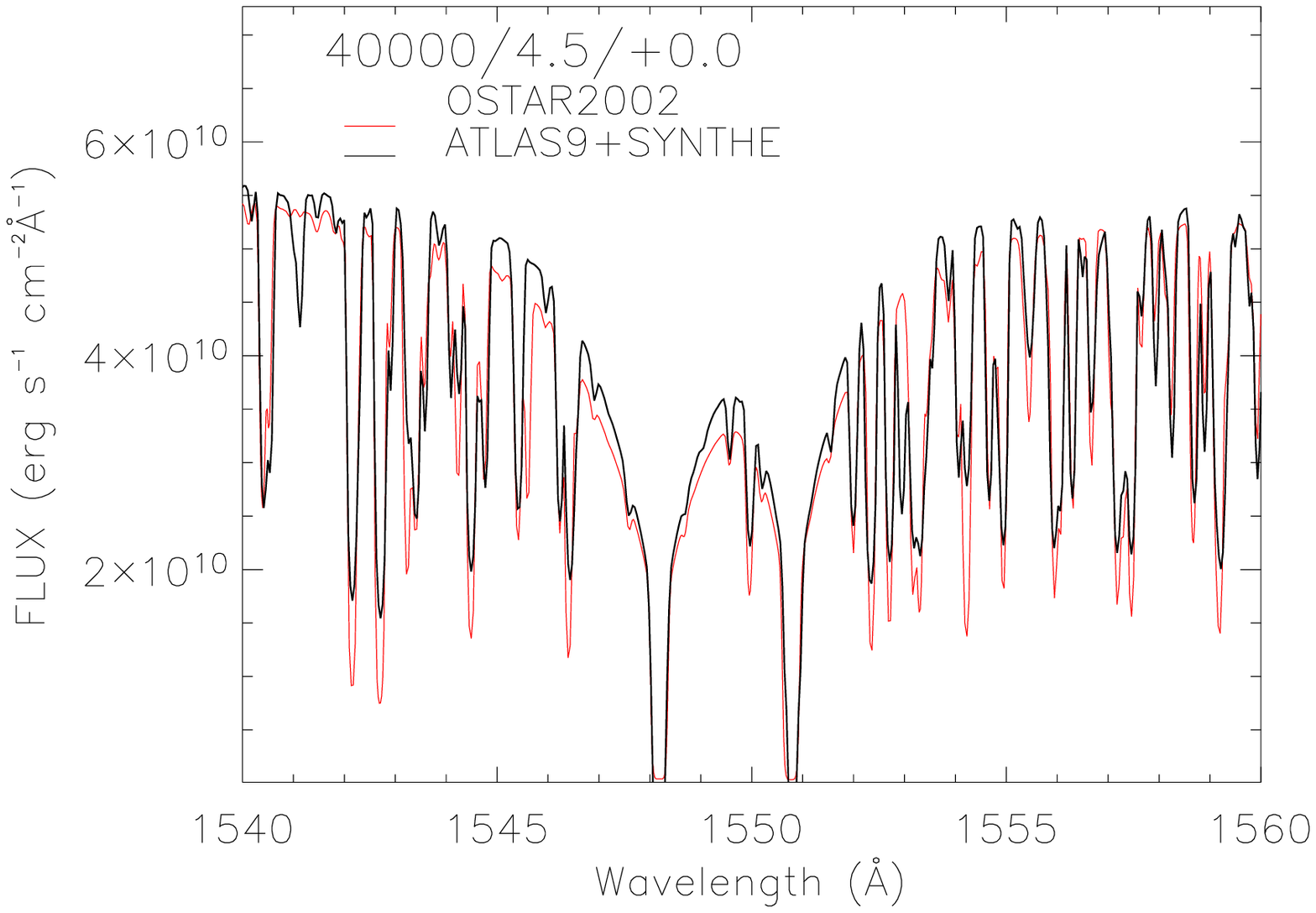,width=0.5\hsize,clip=}
\psfig{file=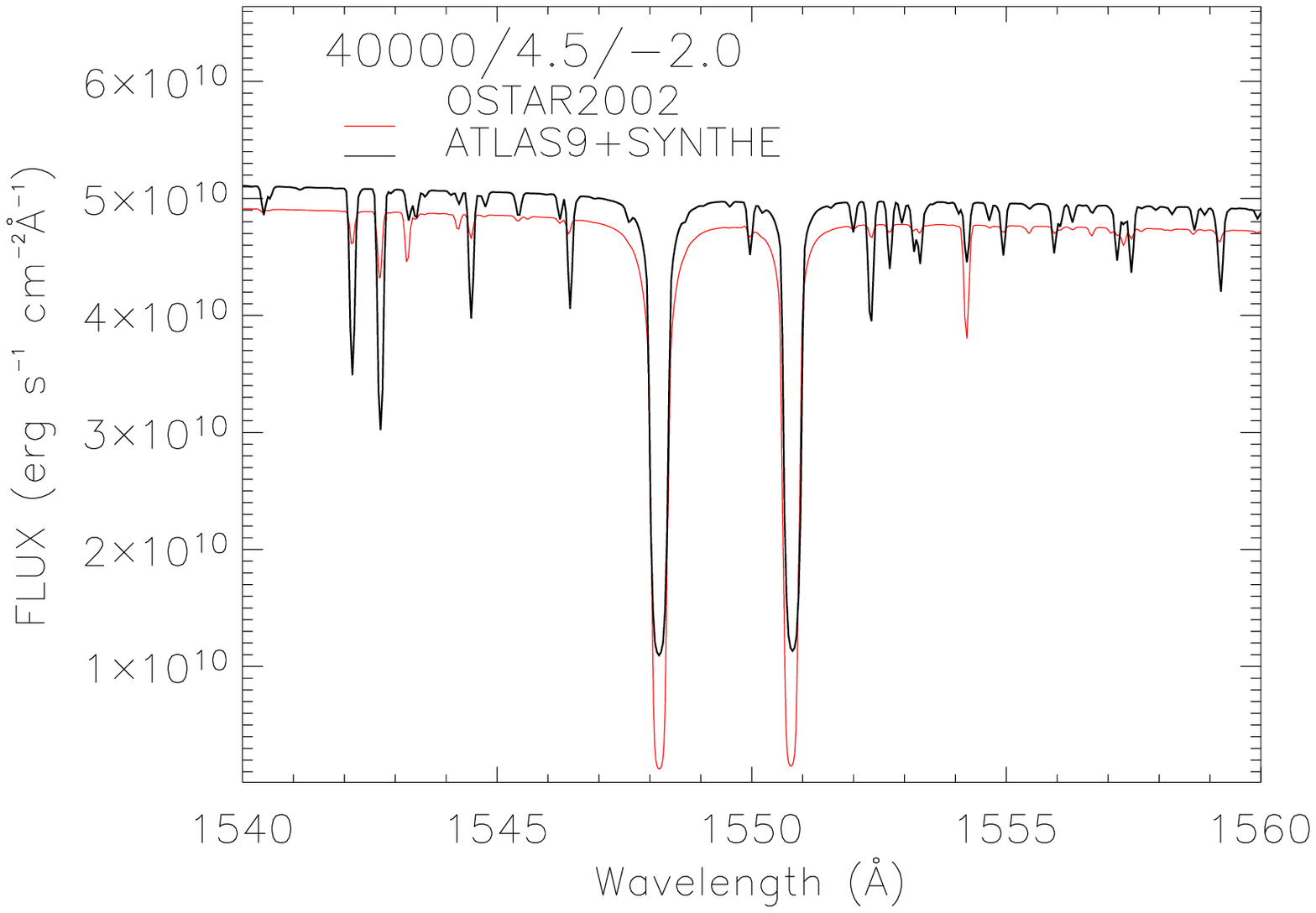,width=0.5\hsize,clip=}
}
\caption{A comparison of LTE and NLTE atmosphere models and synthetic spectra. 
{\it Upper panels:} The upper box displays the temperature profiles of four  
atmosphere models, two models calculated with {\sc Atlas9} and the other two  
from the {\sc Ostar2002} grid. The lower box displays the residuals in the form 
$\Delta T${\sc Tlusty-ATLAS9}. {\it Middle panels:} High-resolution 
($R\sim$ 40\,000) spectra in the wavelength region 1390--1410 \AA, showing two 
prominent lines of Si~\textsc{iv}. {\it Lower panels:} High-resolution 
($R\sim$ 40\,000) spectra in the wavelength region centered on the C~\textsc{iv} 
doublet about 1550 \AA.}
\label{mdlsolar}
\end{figure*}

The temperature structure of {\sc Atlas9} and {\sc Tlusty} model atmospheres 
is compared in the upper panels of Fig.~\ref{mdlsolar}. Although no obvious 
interpretative scheme can be invoked to explain the overall differences in the 
LTE vs.\ NLTE properties, the figure helps assessing some relevant features 
directly dealing with the different temperature stratification in the two sets 
of atmospheres. As a major issue, one sees that, for the same value of \teff, 
the outer structure of the {\sc Tlusty} models strongly modulates with 
metallicity, with a temperature turnover growing at metal-poor abundance; this 
effect can also be recognized in the original models of \citet*{kud76}. 
Typically, temperature inversion begins to appear in \metal$\lesssim -1$ 
models (see Fig.~4 in \citealp{Lanz03}), and involves the $\log \tau_{\sc Ross} 
\lesssim -2$ atmosphere layers. 

On the other hand, when $\tau_{\sc Ross}$ approaches unity, another important 
feature seems to characterize NLTE models. Close to the photosphere region, in 
fact, the NLTE temperature profile appears in general cooler compared with the 
corresponding LTE case. The size of this temperature ``dip'' (cf.\ the 
temperature residuals in Fig.~\ref{mdlsolar}) depends on the detailed physical 
treatment of Helium opacity (see, e.g., the interesting experiments of 
\citealp{kud76}, and \citealp*{werner88}) but, to a more or less pronounced 
extent, it seems to be a quite general feature for NLTE atmospheres. As a 
consequence, for fixed value of \teff, NLTE models tend to be ``redder'' than 
the corresponding LTE solution because the continuum forms in a slightly cooler 
physical environment; this especially reflects in short-wavelength colors, 
like the Str\"omgren $u-b$, as first indicated by \citet{kud76}. Direct 
evidence in the sense of a mildly depressed UV/blue continuum can also be 
derived from the \citet{Lanz03} SEDs, when comparing with the Kurucz LTE models.

The composite trend of temperature stratification leads, in addition, to 
different behaviour of high-resolution features in the SED of stars. As far 
as ``weak'' absorption lines are concerned (that is those mainly generating in 
the innermost regions of star atmospheres), the core emission component induced 
by the temperature turnover of the external NLTE layers should likely act in 
the sense of ``filling'' out the features. This would be especially evident at 
metal-poor regimes, as can be easily recognized in the \metal$= -2$ synthetic 
spectrum of Fig.~\ref{mdlsolar} (right panels). On the contrary, for those 
strong (or saturated) absorption features, whose core forms at the optical 
depth about the region of the temperature ``dip'', we would rather expect a 
slightly increased equivalent width for NLTE models, again, as a result of a 
cooler temperature environment. This is the case, for instance, of some 
Si~\textsc{iv} and C~\textsc{iv} lines (see Fig.~\ref{mdlsolar}, left panel), 
or Hydrogen Balmer absorptions, like $H\gamma$ at 4340~\AA, or the He~\textsc{ii} 
absorption at 4542~\AA, as shown by \citet{kud76}.

The UV SED of late-type stars is also affected by deviations from LTE. In 
addition to the still controversial ``missing solar UV opacity'' 
\citep{Holweger70,Gustafsson75}, in fact, some of the disagreements found when 
comparing the observed and theoretical solar spectrum have been ascribed to 
NLTE effects. However, the recent analysis by  \citet*{Allende03} indicates 
that the claimed discrepancy between theory and observations in the mid-UV 
region around 2400--2600~\AA\ for the Sun cannot be overcome by (and actually 
even worsen with) NLTE calculations.

\subsection{Matching the IUE spectral atlas}

As we have mentioned before, the \citet{wu83} atlas of stellar spectra, 
from the IUE data, still remains a main reference to compare the {\sc Uvblue} 
theoretical output. This composite atlas has been the work-horse of numerous 
investigations on the calibration of stellar features in terms of stellar 
parameters as well as population studies. The original sample of 170 stars has 
been further extended to 220 objects by \citet*{Fan90}, including most late-type 
stars and completing the sampling of spectral types, from O to M, and 
luminosity classes, from I to V. Many subsequent updates have further increased 
the number of entries, and the current Web version of the catalog\footnote{{
\tt http://www-int.stsci.edu/$\sim$jinger/iue.html}} comprises 476 objects.

For the illustrative scope of our work, we restrained the analysis only to the
220 stars of the originally published version of the catalog (hereafter 
referred to as ``the IUE atlas''). Each object of this sample has been 
cross-checked with the literature data, mainly relying on the \citet{Cay97} 
catalog, in order to derive a complete set of atmosphere 
fundamental parameters from high-resolution spectroscopy at optical wavelength. 
In case of multiple determinations for the same star, the average values for 
\teff, \logg, and \metal\ have been considered (retaining, however, only those 
literature sources with complete estimates of {\it all} the three parameters), 
together with a measure of the nominal uncertainty from the data dispersion.

\begin{figure}[!t]
\begin{center}
\psfig{file=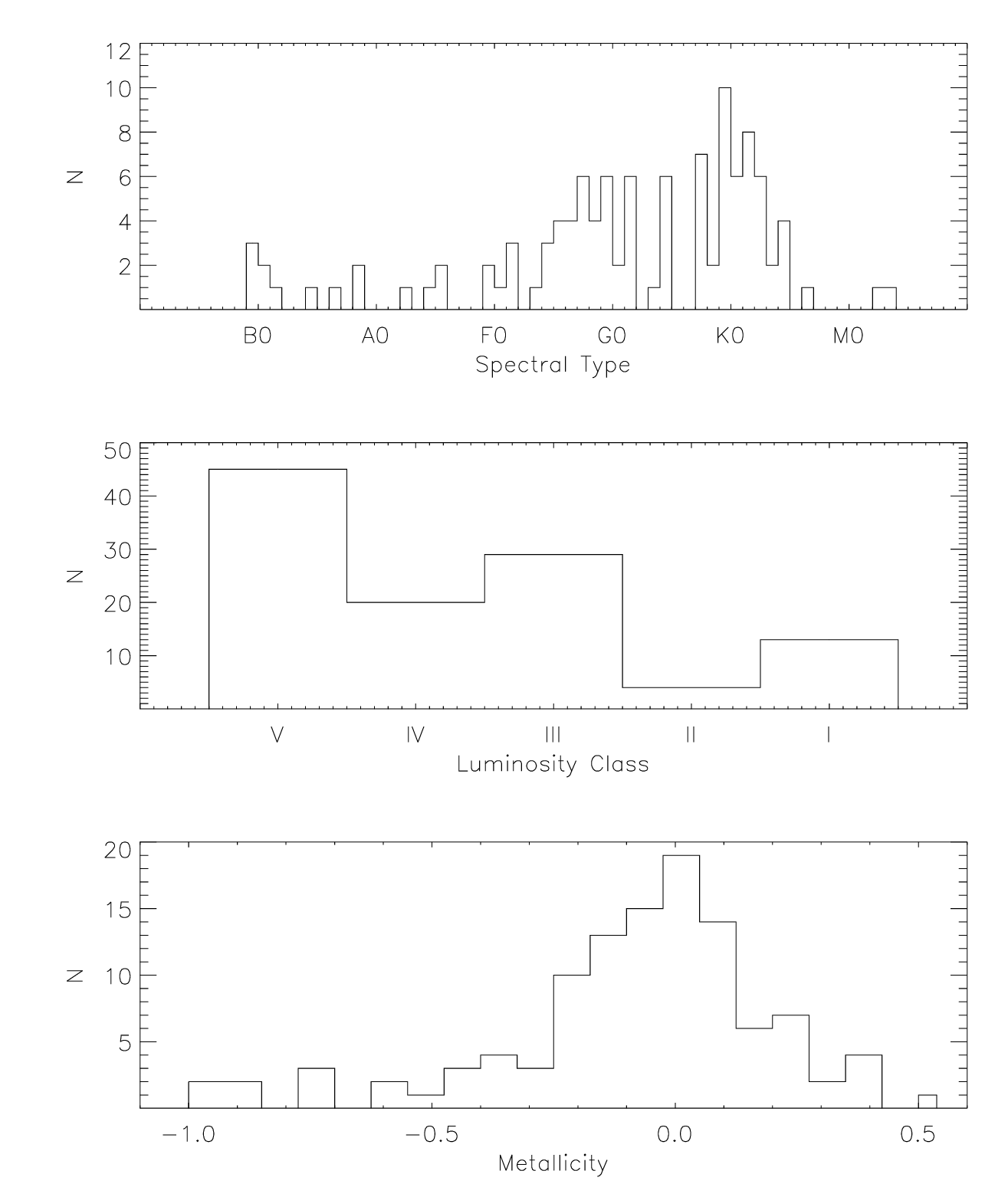,width=\hsize,clip=}
\caption{Distribution of stars in the Wu Atlas as a function of spectral type, 
luminosity and metallicity.}
\label{wustlm}
\end{center}
\end{figure}

Our search procedure eventually secured complete atmosphere parameters for 111 
stars of the original IUE atlas. For this working sample, Table~\ref{iue_atlas} 
collects the relevant information, while the three panels of Fig.~\ref{wustlm} 
report the distribution of stars in spectral type, luminosity class and 
metallicity. In Table~\ref{iue_atlas}, columns 1 to 5 give, respectively, the 
HD number of star, its spectral classification (from the {\sc Simbad} 
database), colour excess (from the \citealp{Fanelli92} compilation), 
atmosphere parameters (from \citealp{Cay97} unless otherwise stated), and 
their rms (for those cases with multiple determinations in the literature).

In order to assure an accurate and homogeneous observational dataset, from the 
updated INES/IUE database\footnote{Our work is based on the INES/IUE general 
archive hosted at the Mexican and Spanish distribution centers of INAOE ({\tt 
http://www.inaoep.mx}) and LAEFF ({\tt http://ines.laeff.esa.es/}), 
respectively.} we extracted and re-processed all the relevant frames for stars 
in our sample. In particular, we considered all the available low-dispersion 
spectra collected in large aperture (20\arcsec) mode and with poor-quality 
flags affecting less than 40\% of the wavelength points.\footnote{Note that 
poor-quality flags quite frequently mark background bad pixels in the original 
IUE frames. Their presence in the extracted 1-D spectra does not necessarily 
indicate a pixel defeat. In any case, visual inspection on each single image 
discerned whether the spectrum was to be rejected or not.} The number of SWP 
and LWP/LWR frames eventually used in our analysis are reported in columns 7 
and 8 of Table~\ref{iue_atlas}.

Wavelength calibration for the whole database was tuned up by fitting several 
prominent features in the spectra according to the stellar type. For example, 
for hot stars we mainly relied on Ly$\alpha$, C~\textsc{iv} at 1550~\AA\ , 
Al~\textsc{ii} at 1617~\AA\ etc., while for cool stars we used the blend of Si 
and Cr lines at 2124~\AA\  and the Mg~\textsc{i} line at 2852~\AA. In any case, 
at least four lines were used for wavelength fine-tuning. For those stars with 
multiple observations, the fiducial SED for each star was obtained by averaging 
the total number of images. 

\begin{figure*}[!ht]
\centerline{
\psfig{file=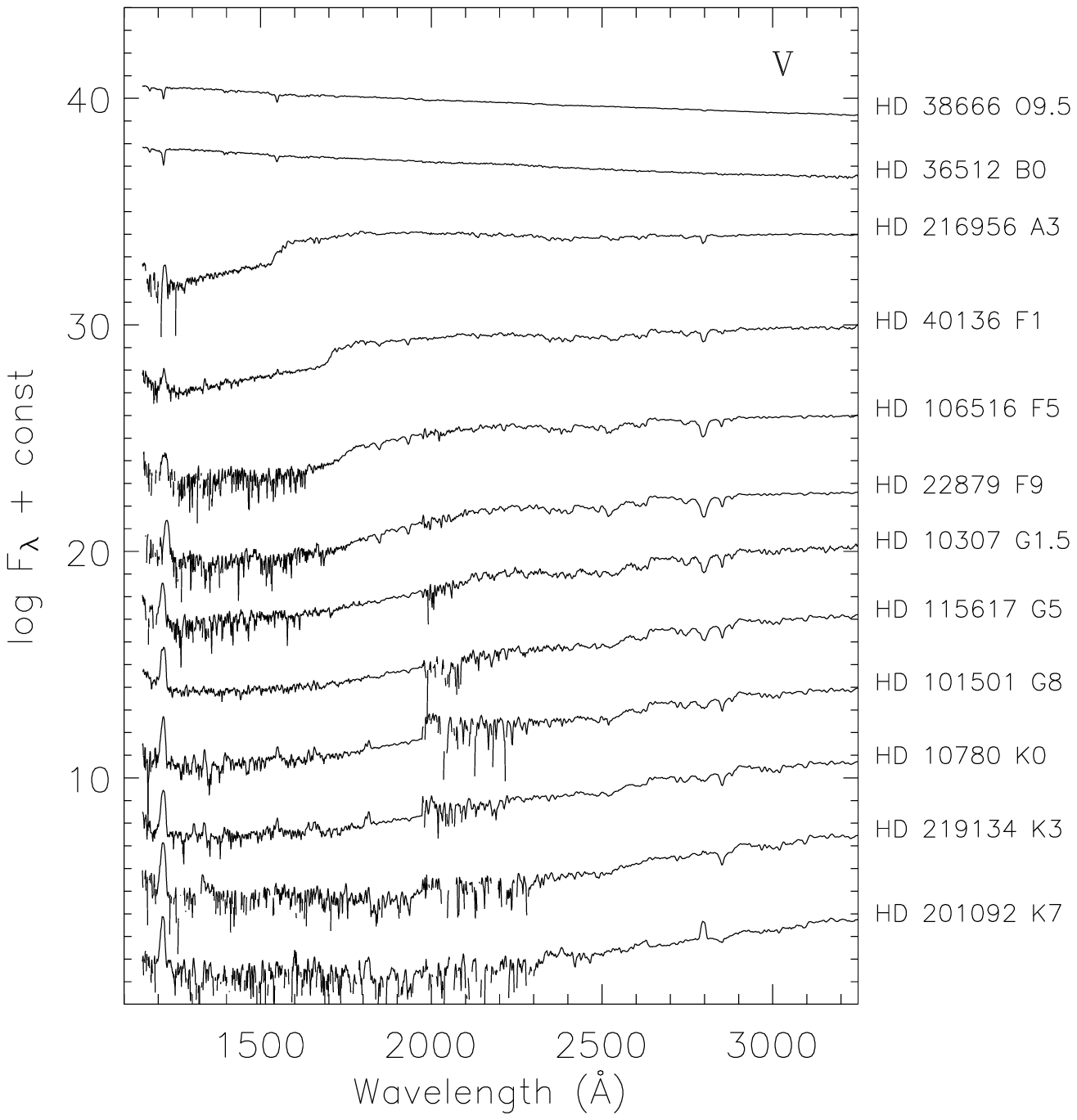,width=0.5\hsize,clip=}
\psfig{file=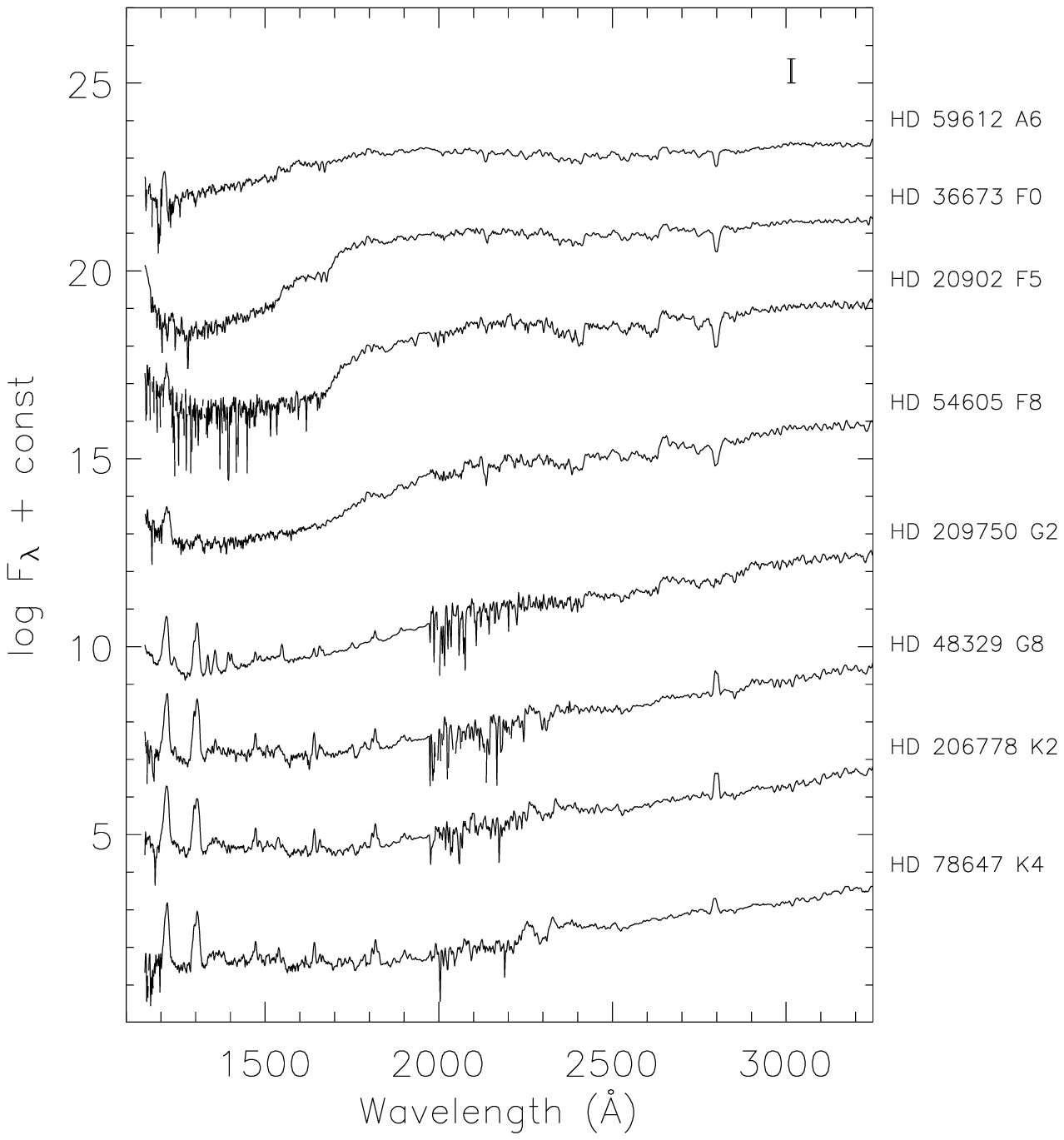,width=0.5\hsize,clip=}
}
\caption{An illustrative selection of SED of the IUE atlas with luminosity 
class V (dwarfs), and I (supergiants).}
\label{wuclassv}
\end{figure*}

The resulting mean spectrum was weighted by the 
errors ascribed to the flux given in the INES files. 
All spectra have then been corrected for Galactic 
reddening according to \citet*{Mathis90}; we assumed $R_{V}$ = 3.1 and the color 
excess E(B-V) as reported by \citet{Fanelli92} (see Table~\ref{iue_atlas}).
Figure~\ref{wuclassv} displays an illustrative example of the SED along the 
spectral-type sequence for stars of different luminosity class.

\begin{figure*}[!ht]
\centerline{
\psfig{file=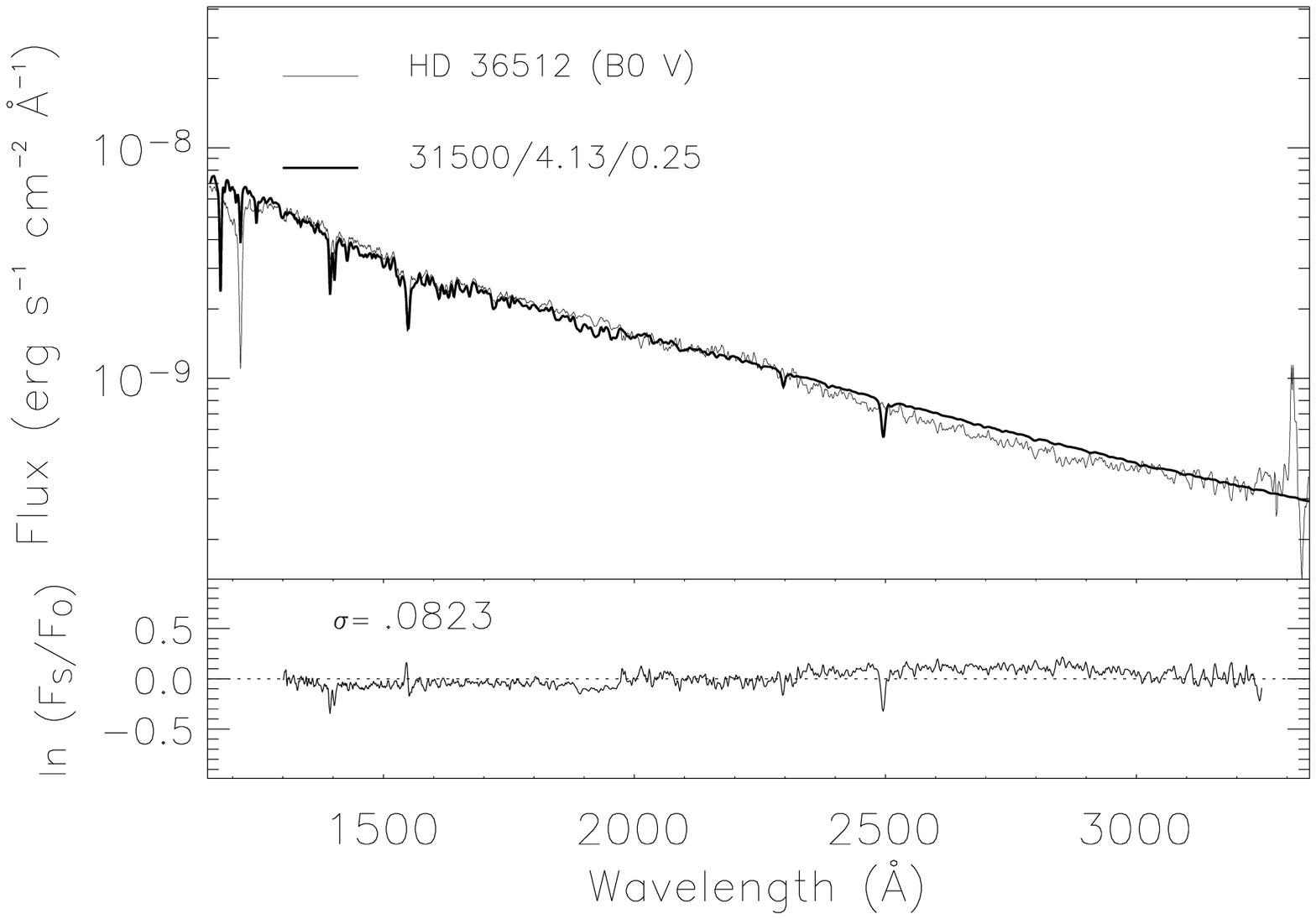,width=0.5\hsize,clip=}
\psfig{file=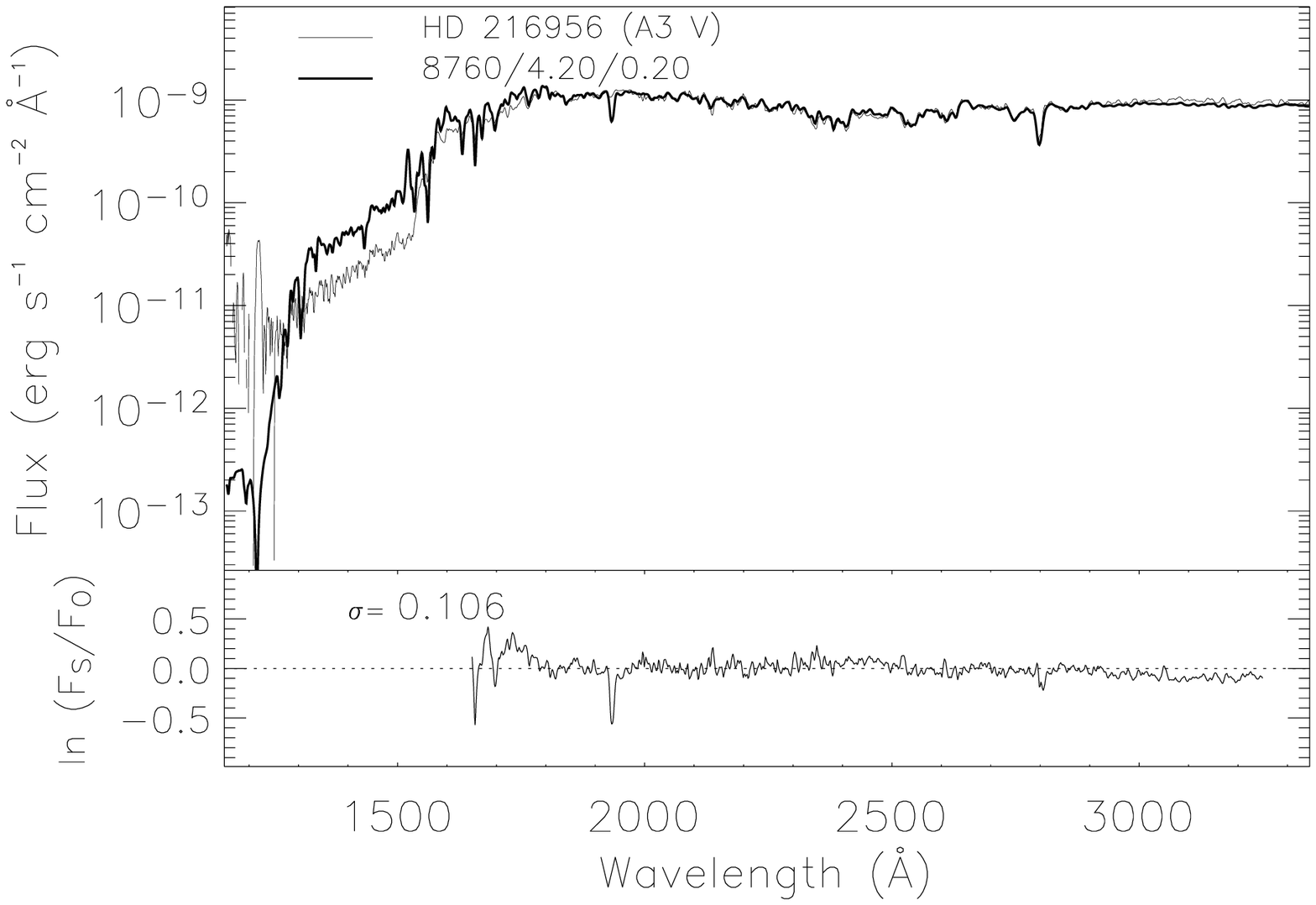,width=0.5\hsize,clip=}
}\centerline{
\psfig{file=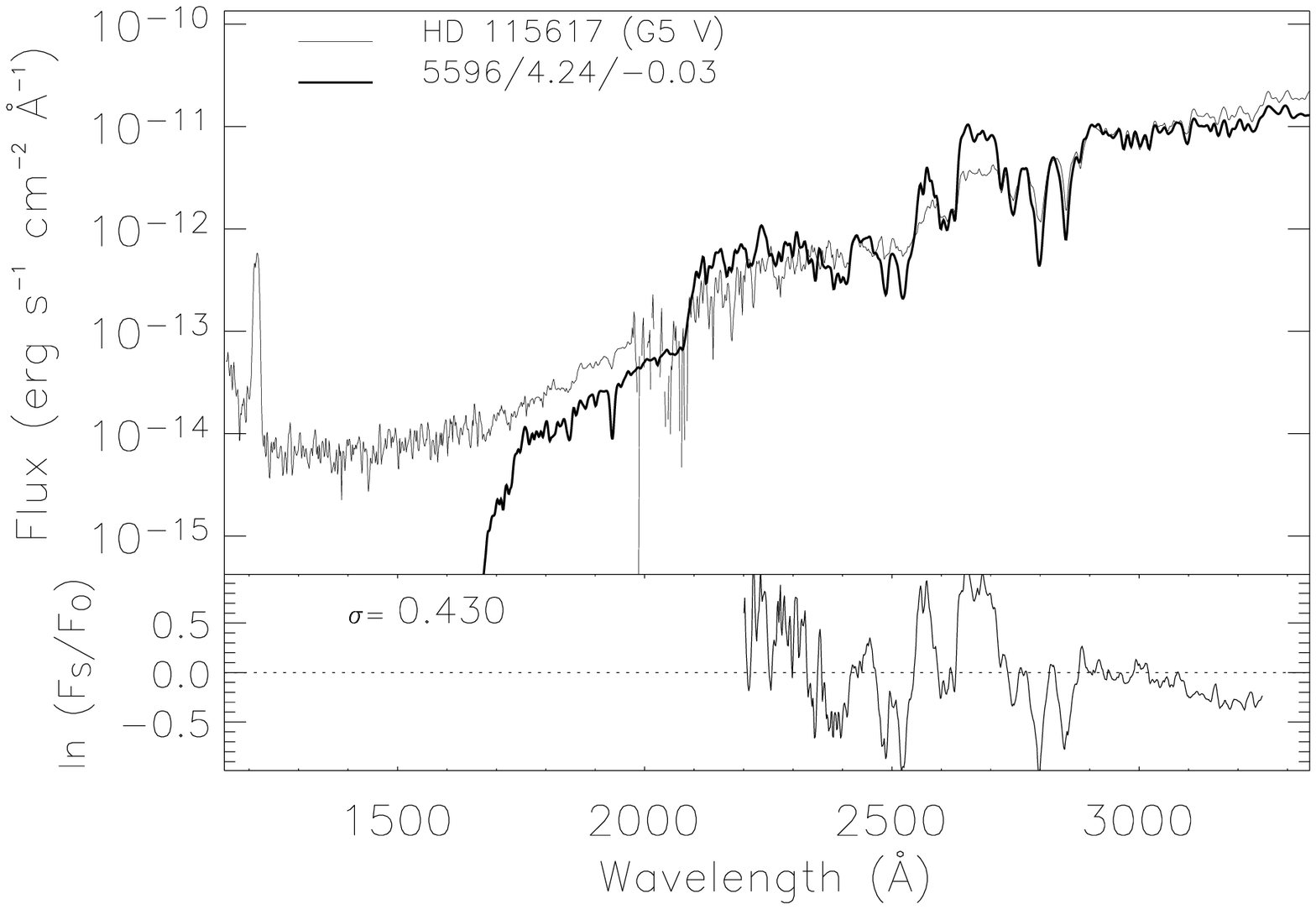,width=0.5\hsize,clip=}
\psfig{file=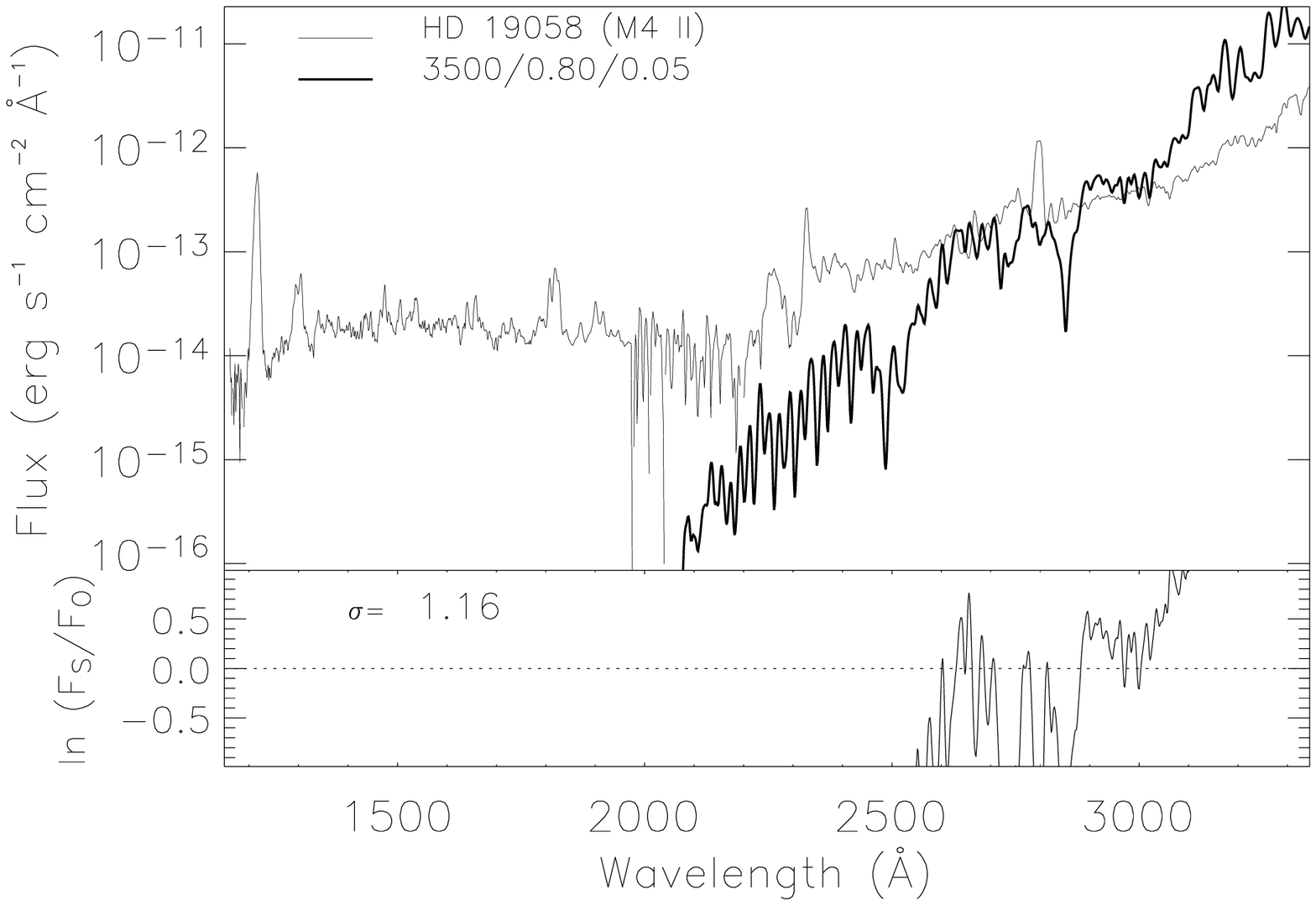,width=0.5\hsize,clip=}
}
\caption{An example of our comparison procedure for a selected set of IUE
stars of different spectral type. Theoretical fluxes have been scaled after a
normalization such that $<\ln f(\lambda)_{\rm IUE} - \ln f(\lambda)_{\rm UVBLUE}>=0$.}
\label{appendix_lino}
\end{figure*}

The observed SED of each star in the IUE working sample has been matched with 
the theoretical spectra by interpolating the {\sc Uvblue} library according to 
the fiducial atmosphere parameters of Table~\ref{iue_atlas}. In order to allow 
a self-consistent comparison, the theoretical SED was then degraded to the IUE 
6~\AA\ wavelength resolution and rebinned at the wavelength points of the 
observations. The comparison was taken over 
a common wavelength range according to the spectral type. For hot stars this 
spanned the 1300-3250~\AA\ interval, excluding the extreme UV region shortward 
of the Ly$\alpha$ line, possibly affected by interstellar absorption 
\citep{ds94,b98}. As emission of cooler stars drops at nominal values in the 
far-UV range, for late spectral types we restrained the SED comparison longward 
of 1650~\AA\ for A-type objects, 2200~\AA\ for the F-G types, and 2500~\AA\ 
for K and M stars. An example of our comparison procedure, for a selected set 
of IUE stars of different spectral type, is shown in Fig.~\ref{appendix_lino}.

A straightforward measure of the ``likeness'' between observed and theoretical 
SED directly derives from the standard deviation of the [({\sc Iue}) - 
({\sc Uvblue})] flux residuals (in the natural logarithm domain), defined as
\begin{equation}
\sigma = \sqrt{Var[\ln f(\lambda)_{\rm IUE} - \ln f(\lambda)_{\rm UVBLUE}]}.
\label{eq:ss}
\end{equation}

This definition is independent of the flux offset between theoretical and 
observed spectra and therefore overcomes any internal uncertainty in the
zero-point calibration.\footnote{Note that our normalization procedure fully 
overcomes any internal uncertainty in the zero-point calibration of the INES/IUE 
database. In this regard, the claimed $\sim$ 7 $\%$ systematic offset of the 
IUE archive with respect to HST/FOS data (e.g. \citet{Gonzalez01}), for instance,
has not effect on our comparison.}

\begin{figure*}[!ht]
\centerline{
\psfig{file=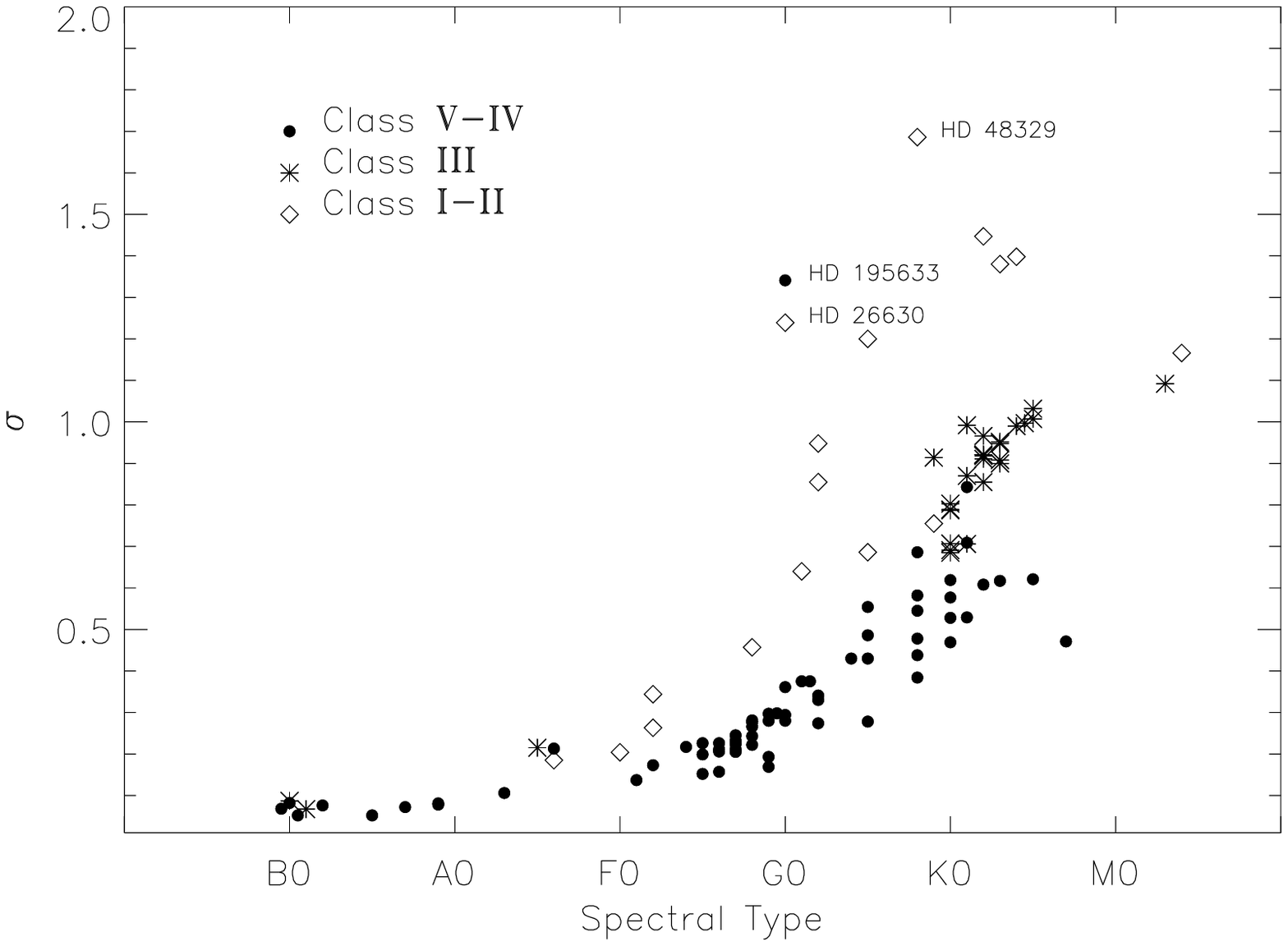,width=0.7\hsize,clip=}
}
\caption{The comparison between observed and theoretical spectra can be 
summarized with this plot. Discrepancies are given by the standard deviation 
as a function of the spectral type ($\sigma$, see text). Three striking 
outliers are labeled in the plot with their HD number (see text for a 
discussion).}
\label{sigma_fit}
\end{figure*}

Figure \ref{sigma_fit} gives an overall summary of $\sigma$ vs.\ spectral type 
and luminosity class for the 111 stars in our working sample; the individual 
values of $\sigma$ for each target star are also reported in column 6 of 
Table~\ref{iue_atlas}. It is evident that the rms value drastically degrades 
for spectral types later than G5, with supergiant stars that are in general 
more poorly reproduced by the {\sc Uvblue} models.

Quite importantly, we point out in this regard that our procedure does 
{\it not} search for any best fit to the IUE observations but simply compares 
the observed spectrum and the corresponding theoretical SED obtained from the 
fiducial atmosphere parameters of the literature. From one hand,  this should 
certainly assure an independent ``acid test'' for the theoretical library, 
though one should also be aware that any inaccurate (or incorrect) set of input 
fundamental parameters (and even more any uncertainty in the reddening correction) for a star 
could reflect in a sensibly increase of the value of $\sigma$.
In addition, one should also consider that the IUE 
atlas actually includes a number of ``peculiar'' objects; a glance to 
column 9 of Table~\ref{iue_atlas} shows that many stars are double or 
multiple systems, with an important fraction of spectroscopic binaries and 
variables. This clearly works in the sense of artificially worsening the match 
with the theoretical models.
We could briefly sketch the relevant results of our comparison across the whole 
spectral-type sequence:
 
\noindent$\bullet$ {\it Early-type stars: O and B}

Most stars in this stellar group have complete SWP+LWP/R spectra and comparison
with the synthetic SED can be carried out over the full wavelength range.
One of the main discrepancies with the {\sc Uvblue} models is the C~\textsc{iv} 
line at 1550~\AA, partly as a consequence of the physical origin of this line,
strongly affected by stellar winds and mass loss processes in massive stars.
Additionally, we also found that the O~\textsc{iii} line at 2496~\AA, predicted 
by the models, is never detected in the observed spectra. 

\noindent$\bullet$ {\it A-type stars}

Although to a lesser extent, also for these stars the C~\textsc{iv} line 
(together with the C~\textsc{i} and Al~\textsc{ii} blend at 1670~\AA) is not 
well matched. As far as low-resolution features are concerned, one main source 
of discrepancy in late A stars is the Si~\textsc{i} spectral break at 1530~\AA,
poorly reproduced by the synthetic spectra. If we neglect the spectral region
around this feature the value of $\sigma$ sensibly improves in at least 70\% of 
the cases. On the other hand, it is also worth noting that, for their 
temperature range, A stars are the strongest far-UV emitters and little 
uncertainty in the estimated value of \teff\ can therefore reflect in a large 
mismatch between observed and theoretical SED (see, e.g.\ \citealp*{Cheng03}, 
for a discussion on FUSE data for star 2 And).

\noindent$\bullet$ {\it Intermediate-type stars: F and G}

Only the region longward of 2200~\AA\ was retained for these stars. Prominent 
metallic features of F stars are always stronger in theoretical flux, in 
particular the Fe~\textsc{ii} blend at 2400~\AA\ and that of Fe~\textsc{i}/
Si~\textsc{i} at about 2500~\AA, as well as the Mg~\textsc{ii} resonance doublet 
at 2800~\AA\ and the Mg~\textsc{i} line at 2852~\AA. The Mg break around 
2600~\AA\ is also poorly reproduced by the models, with the difference 
increasing for decreasing gravity and effective temperature. For G stars, 
chromosphere activity becomes increasingly important, in most cases partially 
filling the Mg~\textsc{ii} doublet with a core emission. In general, the 
2400--2700~\AA\ spectral region remains the most problematic one (see
Fig.~\ref{appendix_lino}, lower left panel) confirming 
the \citet{Allende03} hints for a missing Mg~\textsc{i} opacity, according to 
their mid-UV data of the Sun collected during the UARS satellite mission. 

Curiously enough, all the three striking outliers in Fig.~\ref{sigma_fit} are 
G-type stars, namely:

{\it HD~26630:} a G0 supergiant with enhanced far-UV emission. The star is 
reported in the {\sc Simbad} database as a spectroscopic binary ($P = 283^d.3$) 
and \citet*{Parsons} decomposed the IUE spectrum isolating the contribution of 
a hot companion of spectral type B9.

{\it HD~48329:} according to {\sc Simbad}, this G8 supergiant shows 
microvariability and the HR catalog reports a strong P-Cygni effect for the 
He~10830~\AA\ infrared line. The IUE spectrum shows a moderate UV excess 
compared to the theoretical SED according to the fiducial atmosphere 
parameters, thus suggesting a slightly warmer value for \teff\ or enhanced 
continuum emission due to a chromosphere contribution.

{\it HD~195633:} the UV spectrum of this G0V star is clearly incompatible with 
its spectral type suggesting either a problem with its atmosphere parameters 
or, more likely, with the reddening correction (that in this case might have 
been underestimated).

\noindent$\bullet$ {\it Late-type stars: K and M}

The UV spectra of these stars are dominated by chromospheric emission and only 
the wavelength region redder than $\lambda \gtrsim 2900$~\AA\ is usefully 
reproduced by theoretical models. The Mg~\textsc{ii} doublet is in strong 
emission for stars later than K2III or K5V, while the 2852~\AA\ Mg~\textsc{i} 
photospheric line is far more intense in theoretical spectra, especially for 
giants and supergiants (Fig.~\ref{appendix_lino}, lower right panel).

\section{Summary and future work}

We have built a new grid of theoretical fluxes, the {\sc Uvblue} library, at 
high resolution ($R$= 50\,000) for a wide volume in the parameter space.
The models suitably cover in the H-R diagram the full evolution of high- and 
low-mass stars and consist of nearly 1800 spectra spanning the wavelength 
region of 850--4700~\AA, the most comprehensive ever computed at high 
resolution in the ultraviolet-blue spectral band. Unlike most empirical and 
theoretical libraries, {\sc Uvblue} includes all stellar types (with \teff$\,$ 
comprised between 3000 and 50\,000~K) extending its applicability to the study 
of single stars and stellar systems, through population synthesis models.

Work is in progress to extend the spectral coverage of the grid down to the 
He ionization edge, in the EUV range, and further explore the parameter space 
by including ``Pop III'' stars at \metal$\ll -2$, as well as a wider 
range of microturbulent velocity. The forthcoming implementation in our code 
of H$_{2}$ and Ly$\alpha$ quasi-molecular H-H and H-H$^+$ opacities will also 
allow a substantial refinement in the calculation of far-UV synthetic spectra 
of intermediate-type stars. The recent analysis of \citet{ck01} indicates, in 
fact, that the lack of these opacities might explain the disagreement between 
observed and theoretical spectral energy distributions found in some metal 
deficient A-type stars.
 
In this work we have restricted our analysis just to the presentation of the 
new spectral grid and reviewed some relevant issues for a safe application of 
the theoretical models to UV observations. Our analysis also included a brief 
comparison of our LTE models with the corresponding output from the NLTE 
synthesis code {\sc Tlusty}. As expected, the match at low spectral resolution 
is fairly good (however with NLTE models, in average, slightly ``redder'' 
compared to the LTE SED for the same \teff), while a larger difference could 
be detected for weak lines, that are nearly wiped out by the enhanced core 
emission component in case of NLTE atmospheres. These effects seem to magnify 
at metal-poor abundance (typically \metal$\lesssim -1$).

A general comparison with UV observations has been carried out, mainly relying 
on the subset of stars of the \citet{wu83} and \citet{Fan90} IUE catalog with 
complete atmosphere parameters available from the literature. A match 
with a working sample of 111 objects showed that {\sc Uvblue} models provide
an accurate description of the main mid- and low-resolution spectral features 
for stars along the whole sequence from the B to $\sim$G5 type. The residual 
rms in our comparison procedure, however, sensibly degrades for later spectral 
types, with supergiant stars that are in general more poorly reproduced than 
dwarfs. As a possible explanation of the general trend, we point out that 
observations were compared with theoretical SED built up according to the 
fiducial atmosphere parameters provided by the literature, and {\it no} best 
fit was searched to the observations. This procedure is, obviously, prone to 
any uncertainty in the input set for \teff, \logg$\;$ and \metal, and in the 
reddening correction, both effects naturally reflecting in a poorer value of 
$\sigma$, via eq.~(\ref{eq:ss}). In addition, one should also consider the 
important contamination of the IUE stellar atlas, due to double or multiple 
stars and to variability effects in most of late-type objects. This clearly 
works in the sense of artificially worsening the match between theory and 
observations. 

A parallel effort, aimed at constructing a cleaner and complete sample of 
``normal'' stars from the IUE data, and self-consistently complement the 
estimate of their atmosphere fundamental parameters, is in progress with new 
optical observations collected by our group at the ``G. Haro'' Observatory of 
Cananea (Mexico). A substantially improved agreement with the models should be 
expected for this tuned stellar sample that, by the way, will also provide 
a valuable empirical atlas for reference in future studies.

These results will be the subject of a forthcoming paper of this series (Chavez 
et al., in preparation), where we will also approach in further detail the 
application of {\sc Uvblue} synthetic spectra to the analysis of UV narrow-band 
indices like those of \citet{Fan90}. The further application of the 
{\sc Uvblue} library for high-resolution stellar population synthesis will be 
the natural step of our work (Buzzoni et al., in preparation), relying on 
the \citet{b89,b02} theoretical code to have a deeper look to the main 
distinctive properties of galaxy SED.

\acknowledgments 
Authors are pleased to thank Octavio Cardona at INAOE 
for fruitful discussions and helpful suggestions, that accompanied our work
along the entire {\sc Uvblue} project. We acknowledge financial support from 
Mexican CONACyT, via grant 36547-E, and Italian MURST, under grant COFIN00 
02-016. MC would like to express his gratitude to The Vatican Observatory for 
partially supporting his sabbatical stay at the University of Arizona.

\clearpage


\begin{deluxetable} {rllcccccl}
\tabletypesize{\tiny}
\tablewidth{0pt}
\tablecaption{IUE ultraviolet atlas}
\tablehead{\colhead{HD} & \colhead{Sp. Type} & \colhead{E(B-V)} & 
\colhead{\teff/\logg/\metal} & \colhead{rms} & \colhead{$\sigma$} &
\multicolumn{2}{c}{\# images} & \colhead{Remarks}\\
\colhead{} & \colhead{} & \colhead{} & \colhead{} & \colhead{} &
\colhead{} & \colhead{SWP} & \colhead{LWR/LWP} & \colhead{}
} 
\startdata
   2151&   G2   IV& 0.00& 5793/4.05/-0.17&    71/0.25/0.10&0.34& 14&  3&    Variable \\
   3360&   B2   IV& 0.04&22180/3.92/-0.23&           -/-/-&0.08& 21& 65&    Variable \\
   4128&   K0  III& 0.00&  4836/2.62/0.01&    86/0.22/0.12&0.79&  5&  4&    Variable \\
   4307&   G2    V& 0.00& 5736/3.99/-0.32&   103/0.09/0.06&0.27&  0&  2&           - \\
   4614&   G0    V& 0.00& 5809/4.41/-0.23&   135/0.04/0.06&0.28&  4&  2& Spec-Binary \\
   6203&   K0  III& 0.00& 4560/2.96/-0.35&           -/-/-&0.79&  0&  2&  Dbl-System \\
  10307& G1.5    V& 0.00& 5845/4.35/-0.02&    74/0.05/0.01&0.38&  1&  1&           - \\
  10380&   K3 IIIb& 0.00& 4052/1.43/-0.29&    60/0.47/0.05&0.95&  1&  2&           - \\
  10476&   K1    V& 0.00& 5196/4.50/-0.20&           -/-/-&0.53&  2&  3&    Variable \\
  10700&   G8    V& 0.00& 5229/4.36/-0.53&   125/0.36/0.11&0.38&  8&  3&           - \\
  10780&   K0    V& 0.00&  5419/4.60/0.36&           -/-/-&0.53&  4&  5&           - \\
  14802&   G2    V& 0.00& 5917/4.30/-0.09&    16/0.15/0.13&0.33&  1&  2&           - \\
  17081&   B7   IV& 0.00& 12807/3.64/0.05&   901/0.31/0.46&0.07&  1&  1&           - \\
  17709&   K5  III& 0.00& 3880/1.42/-0.36&           -/-/-&1.01&  1&  1&    Variable \\
  19058&   M4   II& 0.00&  3500/0.80/0.05&           -/-/-&1.17&  4&  7&   Pulsating \\
  19476&   K0  III& 0.00&  4953/3.09/0.10&    24/0.29/0.06&0.69&  0&  1&    Variable \\
  20630&   G5 Vvar& 0.00&  5648/4.41/0.03&    32/0.09/0.04&0.49& 17&  3&    Variable \\
  22049&   K2    V& 0.00& 5058/4.55/-0.20&    79/0.25/0.08&0.61& 78& 15&    Variable \\
  22879&   F9    V& 0.00& 5838/4.15/-0.88&    62/0.17/0.07&0.17&  1&  3&           - \\
  26630&   G0   Ib& 0.19&  5331/1.38/0.02&   159/0.13/0.21&1.24&  7&  2& Spec-Binary \\
  27383&   F9    V& 0.00&  6280/4.55/0.10$^a$&           -/-/-&0.28&  1&  1&  In-Cluster \\
  27561&   F5    V& 0.00&  6700/4.33/0.04$^a$&           -/-/-&0.23&  0&  1&  In-Cluster \\
  27808&   F8    V& 0.00&  6340/4.36/0.10$^a$&           -/-/-&0.27&  2&  2&  In-Cluster \\
  27836&   G1    V& 0.00& 6120/4.65/-0.08$^a$&           -/-/-&0.38&  3&  3&    Variable \\
  28527&   A6   IV& 0.00&  7986/4.18/0.14$^b$&           -/-/-&0.21&  1&  2&    Variable \\
  29139&   K5  III& 0.00& 3903/1.26/-0.10&   150/0.36/0.22&1.03& 20&  1&    Variable \\
  34816& B0.5   IV& 0.00& 30695/4.14/0.38&  1138/0.12/0.88&0.05& 19& 34&           - \\
  35620&   K3IIICN& 0.00& 4238/1.68/-0.14&   131/0.42/0.22&0.95&  0&  2&  Dbl-System \\
  36512&   B0    V& 0.04& 31500/4.13/0.25&           -/-/-&0.08&  2&  3&           - \\
  36673&   F0   Ib& 0.02& 7234/1.38/-0.04&   208/0.33/0.08&0.20&  2&  1&    Variable \\
  37160&   K0 IIIb& 0.00& 4504/2.56/-0.58&   287/0.24/0.17&0.69&  0&  1&           - \\
  38666& O9.5    V& 0.00&31790/4.00/-0.74$^c$&           -/-/-&0.07& 15& 27&    Varaible \\
  38899&   B9   IV& 0.00& 10903/4.00/0.01&    75/0.14/0.09&0.08&  1&  1&  Dbl-System \\
  40136&   F1    V& 0.03& 7127/4.14/-0.13&   250/0.13/0.08&0.14&  3&  3&           - \\
  44478&   M3  III& 0.00&  3600/1.00/0.11&           -/-/-&1.09&  1&  3&    Variable \\
  46328&   B1  III& 0.01&27720/4.00/-0.18&           -/-/-&0.07&  1&  1&    Variable \\
  48329&   G8   Ib& 0.26& 4592/0.81/-0.05&    21/0.17/0.03&1.68&  4&  4&    Variable \\
  49293&   K0 IIIa& 0.00& 4620/2.59/-0.12&           -/-/-&0.80&  0&  2&           - \\
  54605&   F8 Iab:& 0.09&  6222/1.00/0.35&     -/0.28/0.23&0.46&  6&  3&    Variable \\
  54719&   K2  III& 0.00&  4350/2.17/0.02&           -/-/-&0.92&  0&  1&    Variable \\
  55575&   G0    V& 0.00& 5878/4.23/-0.36&   120/0.35/0.11&0.29&  0&  1&           - \\
  59612&   A6Ib/II& 0.14&  8100/1.50/0.08&           -/-/-&0.19&  2&  2&Multiple-Sta \\
  62509&   K0 IIIb& 0.00&  4890/2.66/0.07&    78/0.30/0.27&0.71& 12&  5&    Variable \\
  63922&   B0  III& 0.11& 30300/4.00/0.33&           -/-/-&0.09&  1&  1&  Dbl-System \\
  64606&   G8    V& 0.00& 5139/4.08/-0.99&    68/0.25/0.03&0.44&  0&  1&           - \\
  66141&   K2  III& 0.00& 4250/2.29/-0.36&           -/-/-&0.86&  0&  1&         Dbl \\
  69267&   K4  III& 0.00& 4072/1.73/-0.17&   211/0.14/0.10&0.99&  2&  1&    Variable \\
  70272& K4.5  III& 0.00& 3900/1.59/-0.03&           -/-/-&1.00&  0&  1&    Variable \\
  72184&   K2  III& 0.00& 4525/2.05/-0.05&           -/-/-&0.92&  0&  2&           - \\
  72324&   G9  III& 0.00& 4730/2.08/-0.10&    28/0.25/0.14&0.91&  0&  2&           - \\
  73471&   K1  III& 0.00&  4500/2.36/0.05&           -/-/-&0.99&  0&  1&           - \\
  75732&   G8    V& 0.00&  5196/4.47/0.22&     -/0.06/0.10&0.69&  0&  1&           - \\
  76294&   G9II-II& 0.00& 4870/2.49/-0.01&    26/0.29/0.34&0.75&  2&  2&           - \\
  78647&   K4Ib-II& 0.07&  4235/1.40/0.23&           -/-/-&1.40&  2&  3&    Variable \\
  82328&   F6   IV& 0.00& 6227/3.98/-0.11&   245/0.30/0.10&0.21&  2&  3& Spec-Binary \\
  84441&   G1   II& 0.08& 5343/2.05/-0.08&    66/0.35/0.23&0.64&  0&  2&    Variable \\
  85503&   K2  III& 0.00&  4501/2.37/0.16&   128/0.24/0.19&0.97&  0&  2&           - \\
  89025&   F0  III& 0.00&  7182/2.97/0.12$^d$&           -/-/-&0.16&  1&  1&    Variable \\
  90839&   F8    V& 0.00& 6072/4.41/-0.23&           -/-/-&0.24&  2&  1&  Dbl-System \\
  95272&   K1  III& 0.00& 4494/2.76/-0.08&   407/0.27/0.16&0.87&  1&  1&           - \\
  99028&   F4   IV& 0.06&  6739/3.98/0.06&           -/-/-&0.22&  3&  2& Spec-Binary \\
 101501&   G8    V& 0.00& 5508/4.64/-0.05&    42/0.06/0.12&0.55&  3&  1&    Variable \\
 102870&   F9    V& 0.00&  6124/4.24/0.19&    57/0.10/0.07&0.30&  4&  1&           - \\
 106516&   F5    V& 0.00& 6089/4.25/-0.76&   137/0.24/0.16&0.15&  3&  2&           - \\
 109379&   G5   II& 0.00&  5147/2.15/0.08&    31/0.07/0.27&0.69&  3&  2&    Variable \\
 113139&   F2    V& 0.01&  6890/4.13/0.02$^a$&           -/-/-&0.17&  1&  1&    Multiple \\
 114710& F9.5    V& 0.00&  6008/4.44/0.10&   103/0.06/0.12&0.30&  3&  2&           - \\
 115617&   G5    V& 0.00& 5596/4.24/-0.03&     5/0.25/0.01&0.43&  1&  2&           - \\
 117176&   G4    V& 0.00& 5478/3.75/-0.11&           -/-/-&0.43&  1&  1&           - \\
 125560&   K3  III& 0.00&  4400/2.42/0.00&           -/-/-&0.91&  0&  2&    Variable \\
 126660&   F7    V& 0.00& 6338/4.29/-0.05&           -/-/-&0.25&  3&  1&    Variable \\
 132345&   K3IIICN& 0.00&  4322/2.08/0.05&   105/0.38/0.09&0.90&  0&  1&  Dbl-System \\
 134083&   F5    V& 0.00&  6632/4.50/0.10&           -/-/-&0.20&  2&  2&    Variable \\
 137759&   K2  III& 0.00&  4515/2.67/0.17&    36/0.10/0.19&0.91&  2&  1&    Variable \\
 140573&   K2 IIIb& 0.00&  4555/2.38/0.11&    51/0.44/0.17&0.92&  3&  1&  Dbl-System \\
 142091&   K1  IVa& 0.00& 4800/3.37/-0.04&           -/-/-&0.71&  0&  1&           - \\
 142373&   F8Ve...& 0.00& 5880/4.20/-0.42&    57/0.18/0.08&0.22&  4&  1&           - \\
 142860&   F6   IV& 0.00& 6340/4.10/-0.13$^a$&           -/-/-&0.21&  3&  1&     Varible \\
 142980&   K1   IV& 0.00&  4560/3.22/0.06&           -/-/-&0.84&  0&  2&           - \\
 145328&   K1III-I& 0.00& 4720/3.25/-0.20&           -/-/-&0.71&  0&  1&    Variable \\
 145675&   K0    V& 0.00&  5265/4.49/0.26&   105/0.07/0.08&0.58&  0&  1&           - \\
 147394&   B5   IV& 0.02& 14868/3.81/0.00&    88/0.19/0.48&0.05&  5&  4&    Variable \\
 150680&   G0   IV& 0.00& 5742/3.76/-0.07&    81/0.05/0.12&0.36&  4&  1&        Spec \\
 157244&   K3Ib-II& 0.00&  4582/1.30/0.50&           -/-/-&1.38&  1&  1&           - \\
 159181&   G2 Iab:& 0.15&  5334/1.48/0.21&    56/0.14/0.10&0.95& 13&  3&  Dbl-System \\
 159561&   A5  III& 0.00&  8126/3.97/0.30$^d$&           -/-/-&0.22&  2&  3&    Variable \\
 161471&   F2  Iae& 0.00&  7000/1.30/0.27&           -/-/-&0.34&  2&  2&    Emission \\
 161797&   G5   IV& 0.00&  5463/3.92/0.19&    60/0.18/0.12&0.55&  6&  2&           - \\
 163506&   F2  Ibe& 0.11& 6400/1.20/-0.41&           -/-/-&0.26& 11&  9&   Pulsating \\
 173667&   F6    V& 0.00& 6337/4.08/-0.12&    64/0.25/0.14&0.23&  2&  1&    Variable \\
 182572&   G8IV...& 0.00&  5658/4.15/0.35&   144/0.25/0.14&0.48&  2&  2&    Variable \\
 185144&   K0    V& 0.00& 5169/4.50/-0.24&    37/0.14/0.01&0.47&  3&  4&    Variable \\
 187691&   F8    V& 0.00&  6146/4.34/0.12&     -/0.13/0.02&0.28&  1&  1&           - \\
 188512&   G8   IV& 0.00& 5212/3.09/-0.03&   161/1.03/0.26&0.58&  1&  2&    Variable \\
 193432&   B9   IV& 0.00& 10313/3.75/0.03&   289/0.27/0.21&0.08&  1&  1&  Dbl-System \\
 195633&   G0   Vw& 0.00& 5969/3.75/-1.00&   249/0.07/0.14&1.34&  0&  1&           - \\
 198149&   K0   IV& 0.00& 4965/3.18/-0.18&   123/0.22/0.27&0.62&  2&  1&           - \\
 200580&   F9    V& 0.00& 5727/3.50/-0.75&        -/-/0.07&0.19&  0&  2&    Multiple \\
 201091&   K5    V& 0.00& 4364/4.55/-0.05&    37/0.10/0.04&0.62& 14& 18&    Variable \\
 201092&   K7    V& 0.00& 3868/4.58/-0.19&   133/0.13/0.26&0.47&  2& 14&  Flare-Star \\
 206778&   K2   Ib& 0.20&  4272/1.02/0.02&   129/0.16/0.18&1.45&  5&  2&    Variable \\
 206859&   G5   Ib& 0.17& 4861/1.45/-0.02&    27/0.30/0.04&1.20&  3&  1&    Variable \\
 207978&   F6IV-Vw& 0.00& 6233/4.03/-0.59&   102/0.20/0.07&0.16&  0&  1&           - \\
 209750&   G2   Ib& 0.12&  5251/1.36/0.18&    92/0.12/0.12&0.86& 30&  1&  Dbl-System \\
 215648&   F7    V& 0.00& 6134/4.14/-0.32&   119/0.04/0.05&0.20&  1&  2&           - \\
 216385&   F7   IV& 0.00& 6144/3.93/-0.44&    92/0.13/0.14&0.21&  0&  1&           - \\
 216956&   A3    V& 0.00&  8760/4.20/0.20$^b$&           -/-/-&0.11&  4&  4&    Variable \\
 217877&   F8    V& 0.00& 6000/4.50/-0.10&           -/-/-&0.28&  0&  1&           - \\
 219134&   K3    V& 0.00&  4695/4.48/0.01&    22/0.04/0.14&0.62&  1&  4&  Flare-Star \\
 222368&   F7    V& 0.00& 6106/4.06/-0.31$^a$&   116/0.19/0.13&0.22&  4&  2&    Variable \\
 224930&   G5   Vb& 0.00& 5305/4.49/-0.88&   189/0.13/0.19&0.28&  0&  2& Spec-Binary \\
\enddata
\tablecomments{{Atmosphere parameters from}\\
a- \citet{Gray01}\\
c- \citet{Fitzpatrick99} \\
b- \citet{Gardiner99} \\
d- \citet{Morossi02} \\
}
\label{iue_atlas}
\end{deluxetable}

\clearpage

\end{document}